\documentclass[epj]{svjour}
\usepackage{graphics}
\usepackage{graphicx}
\usepackage[T1]{fontenc}
\usepackage[latin1]{inputenc}
\begin{document}
\title{Spherical Hartree-Fock calculations with linear 
momentum projection before the variation.}
\subtitle{Part II: Spectral functions and spectroscopic factors.}
\author{R.R. Rodr\'{\i}guez--Guzm\'an and K.W. Schmid
}                     
\mail{rayner@tphys.physik.uni-tuebingen.de}
\institute{Institut für Theoretische Physik der Universität Tübingen, Auf der
Morgenstelle 14, D-72076 Tübingen, Germany.}
\date{Received: date / Revised version: date}
%
\abstract{The hole--spectral functions and from these the spectroscopic factors
have been calculated in an Galilei--invariant way for the ground state
wave functions resulting from spherical Hartree--Fock calculations with
projection onto zero total linear momentum before the variation for the
nuclei $^4$He, $^{12}$C, $^{16}$O, $^{28}$Si, $^{32}$S and $^{40}$Ca.
The results are compared to those of the conventional approach which uses
the ground states resulting from usual spherical Hartree--Fock calculations
subtracting the kinetic energy of the center of mass motion before the
variation and to the results obtained analytically with oscillator occupations.
\medskip\noindent
} 
\authorrunning{R.R. Rodr\'{\i}guez--Guzm\'an and K.W. Schmid}
\titlerunning{Spherical Hartree-Fock calculations with linear  momentum ...}
\maketitle

\section{Introduction}
In the first \cite{ref1.} of the present series of two papers we have demonstrated
that in the nuclear many--body problem Galilei--invariance can be restored
with the help of projection techniques not only for simple oscillator
configurations as they have been used in the recently published analytical
model investigations \cite{ref2.,ref3.,ref4.,ref5.}, but also for more realistic wave functions.
For this purpose, spherical Hartree--Fock calculations with projection into
the center of mass rest frame before the variation have been performed for the
six nuclei $^4$He, $^{12}$C, $^{16}$O, $^{28}$Si, $^{32}$S and $^{40}$Ca. The
results have been compared with those of conventional spherical Hartree--Fock
calculations corrected for the center of mass motion by subtracting its
kinetic energy from the hamiltonian before or after the variation (and thus
already taking the trivial 1/A effect into account). As
single particle basis in all nuclei up to 19 oscillator major shells have
been included, and as effective interaction the Brink--Boeker force B1 \cite{ref6.}
complemented with a short range two--body spin--orbit term derived from
the parametrisation D1S \cite{ref7.} of the Gogny--force has been taken. The
results were also compared to the analytical ones obtained with the same
hamiltonian for simple oscillator determinants in ref. \cite{ref4.}.

For the above mentioned nuclei the oscillator ground states are all
``non--spurious'', i.e. they contain no center of mass excitations.
Consequently, the projected and corrected approaches yield here the same
total binding energy. This is not the case in the Hartree--Fock prescription.
It was shown that the energy gains of the Galilei--invariance conserving
projected calculations with respect to the only corrected ones amount in all
these nuclei to a considerable portion of the energy gains due to major shell
mixing in the latter and are hence equally important.

Drastic effects of the restoration of Galilei--invariance have been obtained
for the hole--energies in the above nuclei, too. As already observed
for the oscillator determinants in ref. [4], also in the Hartree--Fock 
prescription the holes out of the last occupied shell remain almost
unaffected while for those out of the second and third but last occupied
shells the projected energies are considerably different from their
conventionally corrected counterparts (which obviously already include
the trivial 1/A effect).

\begin{figure*}
\begin{center}
\includegraphics[angle=-90,width=12cm]{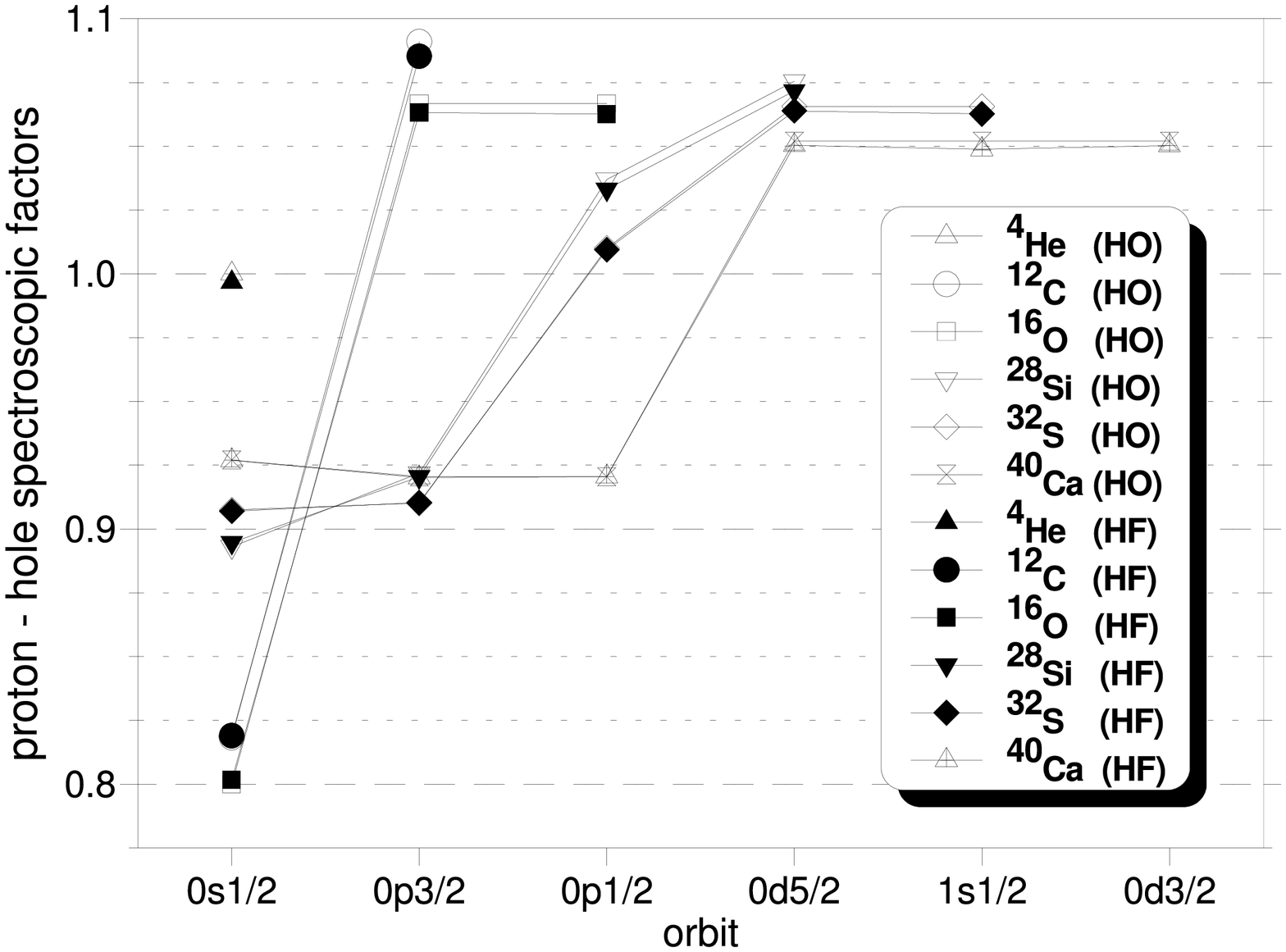}
\caption{
The proton--hole--spectroscopic factors for the various
spherical hole--orbits in the considered nuclei are displayed. Open
symbols refer to the projected results $S^{proj; osc}_{\tilde h}$ out of
eq. (26) using pure harmonic oscillator occupations. In this case
for $^{32}$S and $^{40}$Ca ``0s1/2'' denotes the (with respect to the
1s1/2--orbit) orthonormalized states. Full symbols refer to the general
results out of eq. (25) based on the Hartree--Fock determinants
obtained in ref. [1] with projection into the center of mass rest frame
before the variation. In this case obviously ``0s1/2'' denotes the lowest
s1/2 solution resulting from eq. (18), ``1s1/2'' the second lowest one,
and for the other orbits ``0lj'' always the lowest solution is meant.
Note, that in the usual approach all the displayed numbers should be identical
to one, irrespective whether pure oscillator-- or unprojected Hartree--Fock
determinants are considered.
}
\end{center} 
\end{figure*}

Furthermore, in ref. \cite{ref1.} the elastic charge form factors and corresponding
charge densities as well as the mathematical Coulomb sum rules have been
analyzed. Here again, for oscillator configurations the conventional approach
complemented with the usual Tassie--Barker correction \cite{ref8.} and the projection
yield identical results. In contrast to the results for the total
binding energies, this is at least approximately true in the Hartree--Fock
prescription, too. It should be stressed, however, that form factors can be
rather sensitive to the particular effective interaction used in the
calculations (as it has been discussed already in ref. \cite{ref9.}), and hence this
agreement may dissapear, if more realistic interactions are studied.

\begin{figure*}
\begin{center}
\includegraphics[angle=-90,width=12cm]{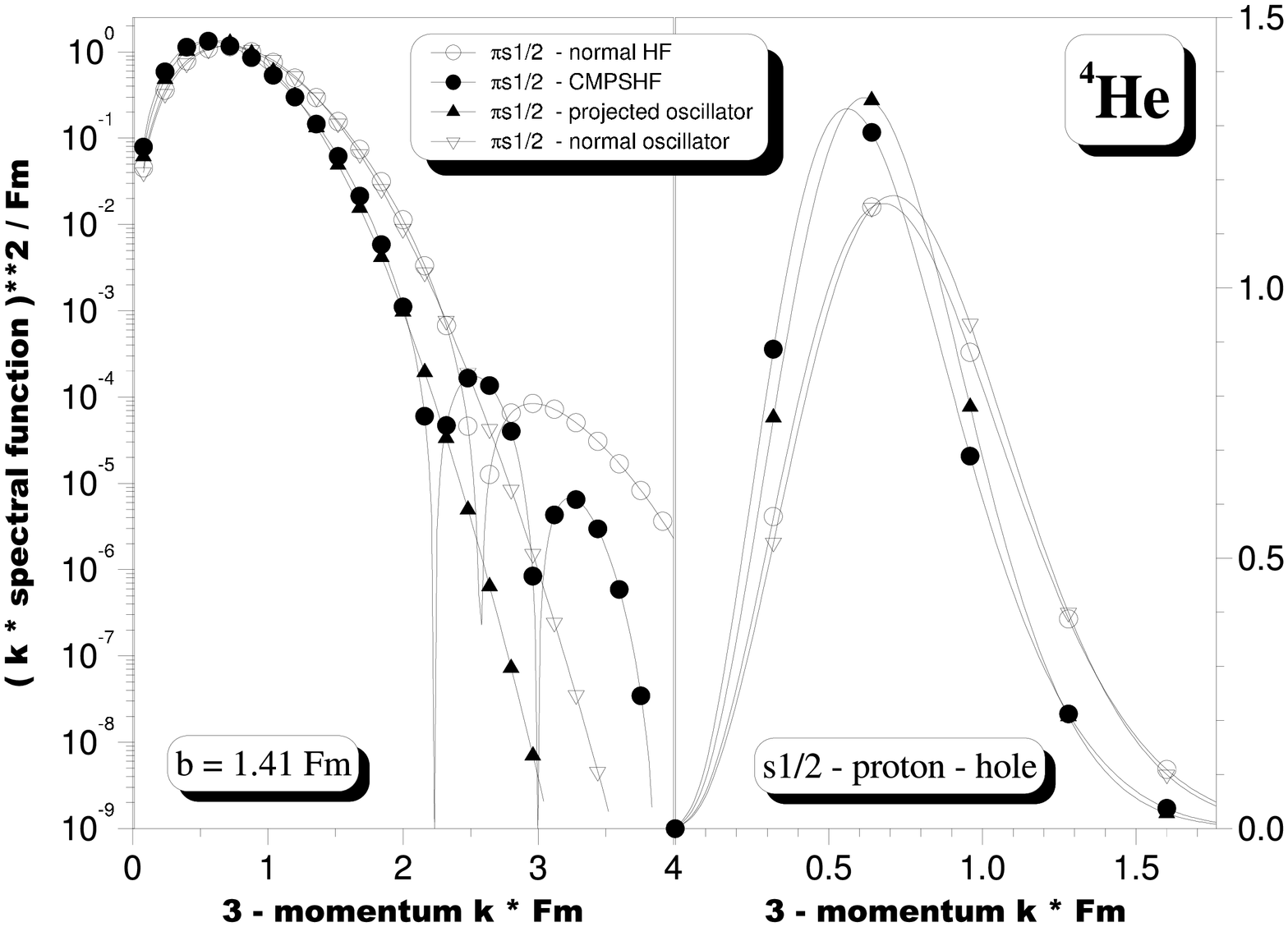}
\caption{
The square of 3--momentum k times the square of the (reduced)
spectral function is plotted versus the 3--momentum for the s1/2--proton--hole
in the nucleus $^4$He. Open circles refer to the results (7) of normal 
Hartree--Fock with subtraction of the kinetic energy of the center of mass
motion before the variation, full circles to the results (22) of Hartree--Fock
with projection into the center of mass rest frame before the variation.
For comparison the normal (open inverted triangles) and projected (full
triangles) for pure oscillator occupation are presented. In the left part
of the figure we display the corresponding curves in a logarithmic, in the
right part in a linear scale. The integral over 3--momentum from zero to
infinty yields for the ``normal'' approaches just one, while for the
projected calculations the spectroscopic factors out of figure 1 are
obtained. The oscillator length was here $b=1.41$ Fm.
}
\end{center} 
\end{figure*}

We shall now continue the analysis of the various wave functions obtained in
ref. \cite{ref1.} by investigating hole--spectral functions and the corresponding
spectroscopic factors. They play an important role in the analysis of
e.g., quasi--elastic electron scattering and one--nucleon
transfer reactions, where they are often used to draw conclusions on
nucleon--nucleon correlations in the considered nuclei. Such conclusions
obviously require the precise knowledge on how ``uncorrelated'' systems
do behave. Now, in the investigation \cite{ref2.} with oscillator determinants
it was already demonstrated that the conventional picture of an
uncorrelated system has to be modified : instead of the usual
spectroscopic factors of one for all the occupied orbits,
Galilei--invariance requires a considerable depletion of the spectroscopic
factors for hole--states with excitation energies larger or equal to
$1\hbar\omega$, while in the last occupied shell an enhancement of the
spectroscopic factors is obtained. Consequently, in the analysis of
correlations not the usual, but the Galilei--invariance respecting
projected spectroscopic factors should be taken as reference. However, 
in ref. \cite{ref2.} only simple oscillator configurations were studied, and it is
not clear a priori whether the much more realistic Hartree--Fock states
produce similar features. This question will be answered in the present
paper.

\section{Spectral functions and spectroscopic factors.}
In the following we shall first summarize the usual definition of 
hole--spectral functions and spectroscopic factors and then present their
Galilei--invariant form. Since we want to evaluate them with the
Galilei--invariant Hartree--Fock ground states obtained in ref. \cite{ref1.} we shall
restrict ourselves in the derivation to ``uncorrelated'' systems, i.e. to
one--determinant states of the form
\begin{equation} \label{Eq1}
\vert D\rangle\,\equiv\,\prod\limits_{h=1}^A\,b_h^{\dagger}\vert 0\rangle,
\end{equation}
where $\vert 0\rangle$ denotes the particle vacuum. The determinant (1)
is composed out of single particle states
\begin{equation} \label{Eq2}
\vert h\rangle\,=\,b_h^{\dagger}\vert 0\rangle\,=\,
\sum\limits_{i=1}^{M_b}\vert i\rangle D_{ih}^*\,=\,
\sum\limits_{i=1}^{M_b} c_i^{\dagger}\vert 0\rangle D_{ih}^* 
\end{equation}
which are obtained by a unitary, in general $M_b\times M_b$--dimensional,
transformation $D$ from the $M_b$ spherical basis states
$\vert i\rangle\,=\,\vert \tau_i n_i l_i j_i m_i\rangle$ defining our model
space. The corresponding creation operators will be denoted by
$\{c_i^{\dagger},\,i=1,...,M_b\}$. For these basis states we shall take
spin--orbit coupled spherical harmonic oscillator wave functions. As usual
$\tau_i$ denotes the isospin projection, $n_i\,(=0,1,...)$ the node number,
$l_i$ the orbital angular momentum, which is coupled with the spin to total
angular momentum $j_i$, and $m_i$ is the 3--projection of the latter. We shall
furthermore restrict ourselves to transformations (\ref{Eq2}), which conserve the
spherical symmetry of the basis orbits. Then each hole state $\vert h\rangle$
has definite $\tau_h$, $l_h$, $j_h$ and $m_h$ and the transformation (2) does
not depend on $m_h$. For each set of quantum numbers the sum runs only over
the node number $n$ and eq. (\ref{Eq2}) reduces to
\begin{equation} \label{Eq3}
\vert h\rangle\,=\,\vert \tau_h \alpha_h l_h j_h m_h\rangle\,=\,
\sum\limits_{n}
\vert \tau_h n l_h j_h m_h\rangle\, D^{\tau_h l_h j_h}_{n \alpha_h},
\end{equation}
where we have assumed in addition that the transformation
matrix $D$ is purely real. For doubly--even nuclei with closed j--shells
the determinant (\ref{Eq1}) is then spherically symmetric, too, and has total
angular momentum and parity $0^+$.

\begin{figure*}
\begin{center}
\includegraphics[angle=-90,width=12cm]{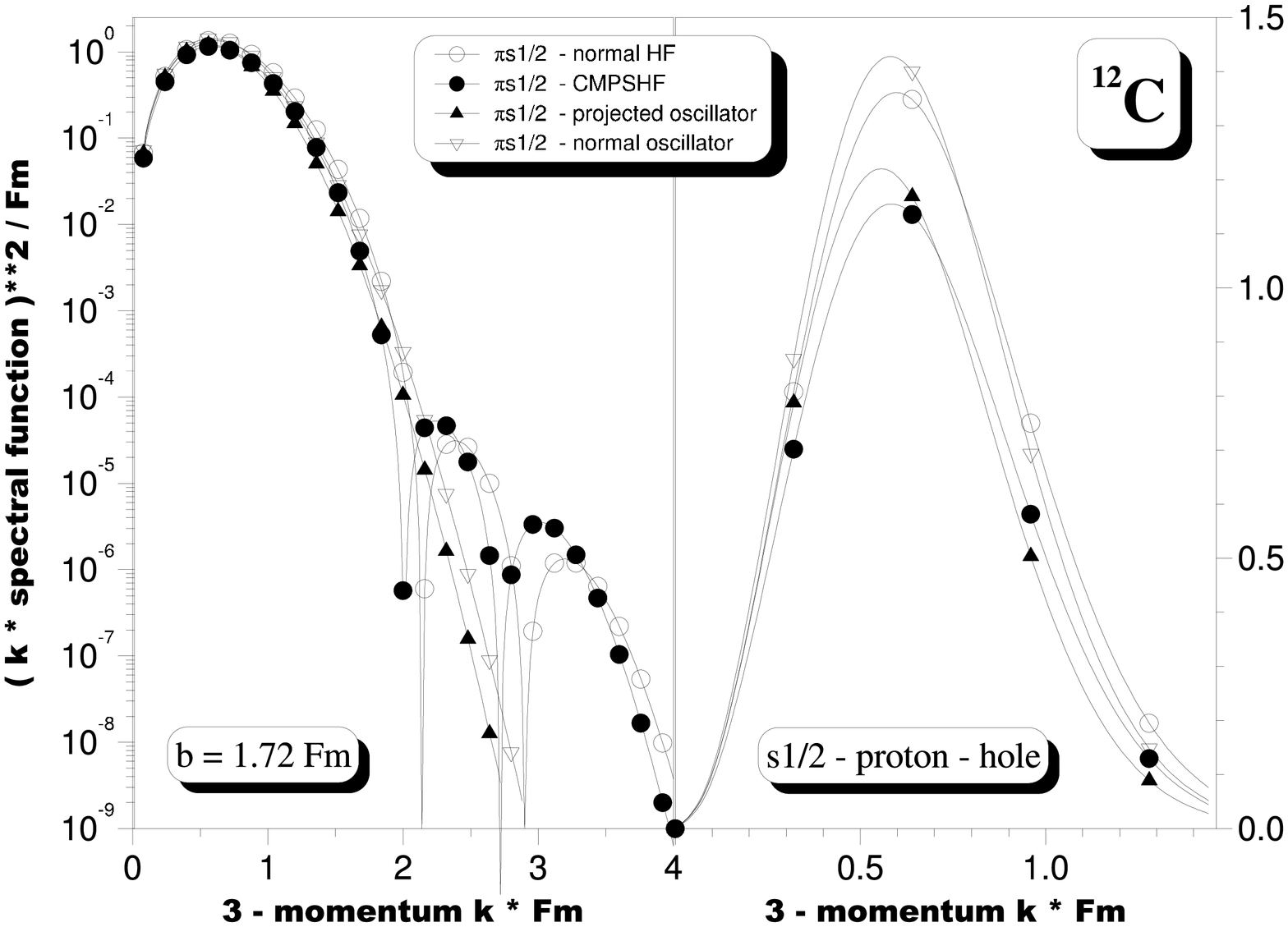}
\caption{
Same as in Fig. 2, but for the s1/2--proton--hole in the nucleus
$^{12}$C. Here the oscillator length was $b=1.72$ Fm.
}
\end{center} 
\end{figure*}

Furthermore we shall introduce creation operators $C_{\vec k\,\tau\sigma}$
\begin{equation} \label{Eq4}
\vert\vec k\,\tau\sigma\rangle\,\equiv\,C_{\vec k\,\tau\sigma}\vert 0\rangle,
\end{equation}
which create from the particle vacuum a nucleon in a plane wave state
with linear momentum $\hbar\vec k$ and spin-- and isospin--projections
$\sigma$ and $\tau$. As usual, the hole--spectral function is then defined
as the probability amplitude to pick out such a plane wave nucleon from
a definite hole state $h$
\begin{equation} \label{Eq5}
f_{h\tau\sigma}^{nor}(\vec k\,)\,\equiv\,\langle D\vert 
C_{\vec k\,\tau\sigma}^{\dagger} b_h\vert D\rangle
\,=\,\langle h\vert\vec k\,\tau\sigma\rangle
\end{equation}
where the superscript $nor$ indicates that the usual (or ``normal'')
prescription is used. For all the hole states $h$ occupied in the 
one--determinant state (\ref{Eq1}) this expression gives essentially 
(the complex conjugate of) the Fourier--transform of the corresponding
single particle wave function. Evaluation of eq. (\ref{Eq5}) yields right away
\begin{eqnarray} \label{Eq6}
f_{h\tau\sigma}^{nor}(\vec k\,)\,=\,\delta_{\tau\tau_h}\,
i^{l_h}\,\sum\limits_{\lambda_h}\,Y_{l_h\lambda_h}^*(\hat k) \times
\nonumber\\
\times
(l_h 1/2 j_h\vert \lambda_h\sigma m_h)\,g_{\tau_h\alpha_h l_h j_h}^{nor}(k),
\end{eqnarray}
where the ``reduced'' spectral function $g_{\tau_h\alpha_h l_h j_h}^{nor}(k)$
is given by
\begin{equation} \label{Eq7}
g_{\tau_h\alpha_h l_h j_h}^{nor}(k)\,=\,
\sum\limits_{n}\,(-)^n\,D^{\tau_h l_h j_h}_{n\alpha_h}\,R_{nl_h}(k)
\end{equation}
and
\begin{eqnarray} \label{Eq8}
R_{nl_h}(k) = b^{3/2}\exp\left\{-{1\over 2}(bk)^2\right\}
{\tilde R}_{nl_h}(bk)
\nonumber\\
=(-)^n \sqrt{{2\over{\pi}}}\int\limits_0^{\infty} dr r^2
j_{l_h}(kr)\, R_{nl_h}(r)
\end{eqnarray}
is the Fourier--transform of the usual radial harmonic oscillator function
$R_{nl_h}(r)$.

\begin{figure*}
\begin{center}
\includegraphics[angle=-90,width=12cm]{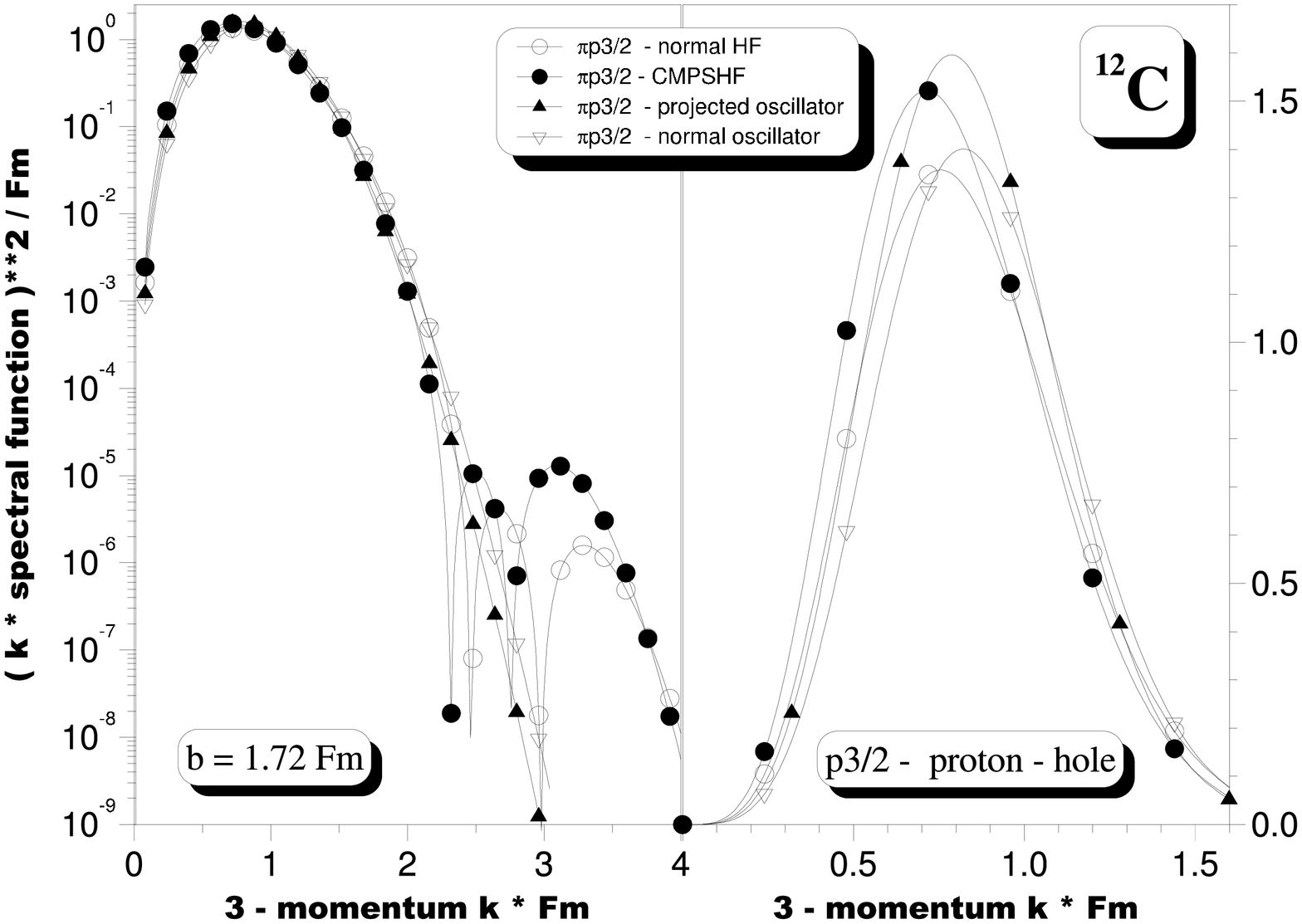}
\caption{
Same as in Fig. 3, but for the p3/2--proton--hole in the nucleus
$^{12}$C.
}
\end{center} 
\end{figure*}

The ``normal'' hole--spectroscopic factors are then defined as
\begin{eqnarray} \label{Eq9}
S_h^{nor}= \sum\limits_{\sigma}\int d^3\vec k\,
\vert f_{h\tau\sigma}^{nor}(\vec k\,)
\vert^2
\nonumber\\
=\delta_{\tau\tau_h}\int\limits_0^{\infty} dk k^2\,
{g_{\tau_h\alpha_h l_h j_h}^{nor}(k)\,}^2\,=\,\delta_{\tau\tau_h}.
\end{eqnarray}
Since the plane waves (\ref{Eq4}) form a complete set, it is obvious that the
hole--spectroscopic factors fulfill the sum rule
\begin{equation} \label{Eq10}
\sum\limits_{h}S_h^{nor}\,=\,A.
\end{equation}
Eqs. (\ref{Eq1}) to (\ref{Eq10}) summarize the usual picture of an uncorrelated system :
the hole--spectroscopic factors are equal to one for all occupied states and
vanish for the unoccupied ones, and the hole--spectral functions are nothing
but the wave functions of the occupied single nucleon states in momentum
representation.

\begin{figure*}
\begin{center}
\includegraphics[angle=-90,width=12cm]{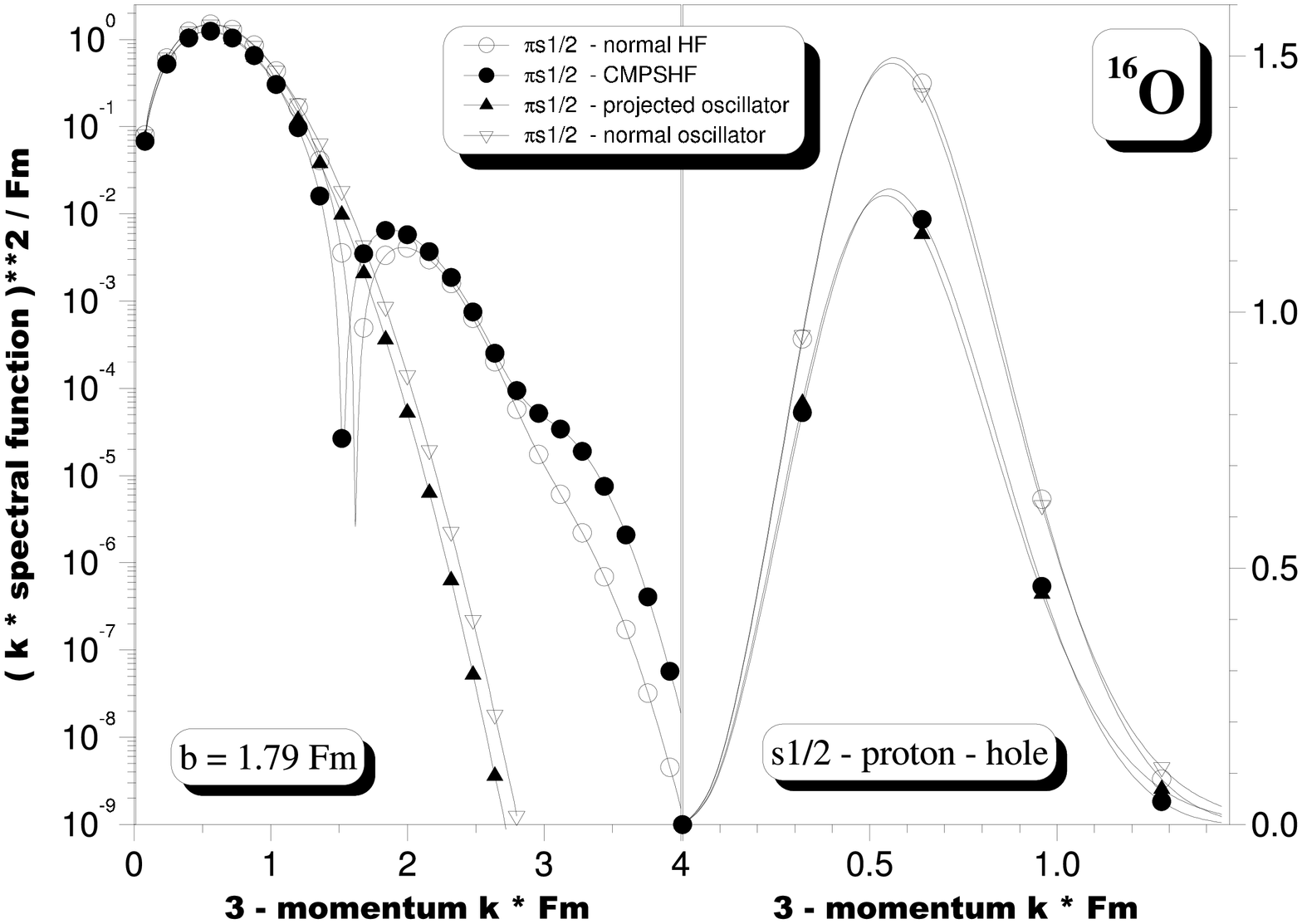}
\caption{
Same as in Fig. 2, but for the s1/2--proton--hole in the nucleus
$^{16}$O. Here the oscillator length was $b=1.79$ Fm.
}
\end{center} 
\end{figure*}

\begin{figure*}
\begin{center}
\includegraphics[angle=-90,width=12cm]{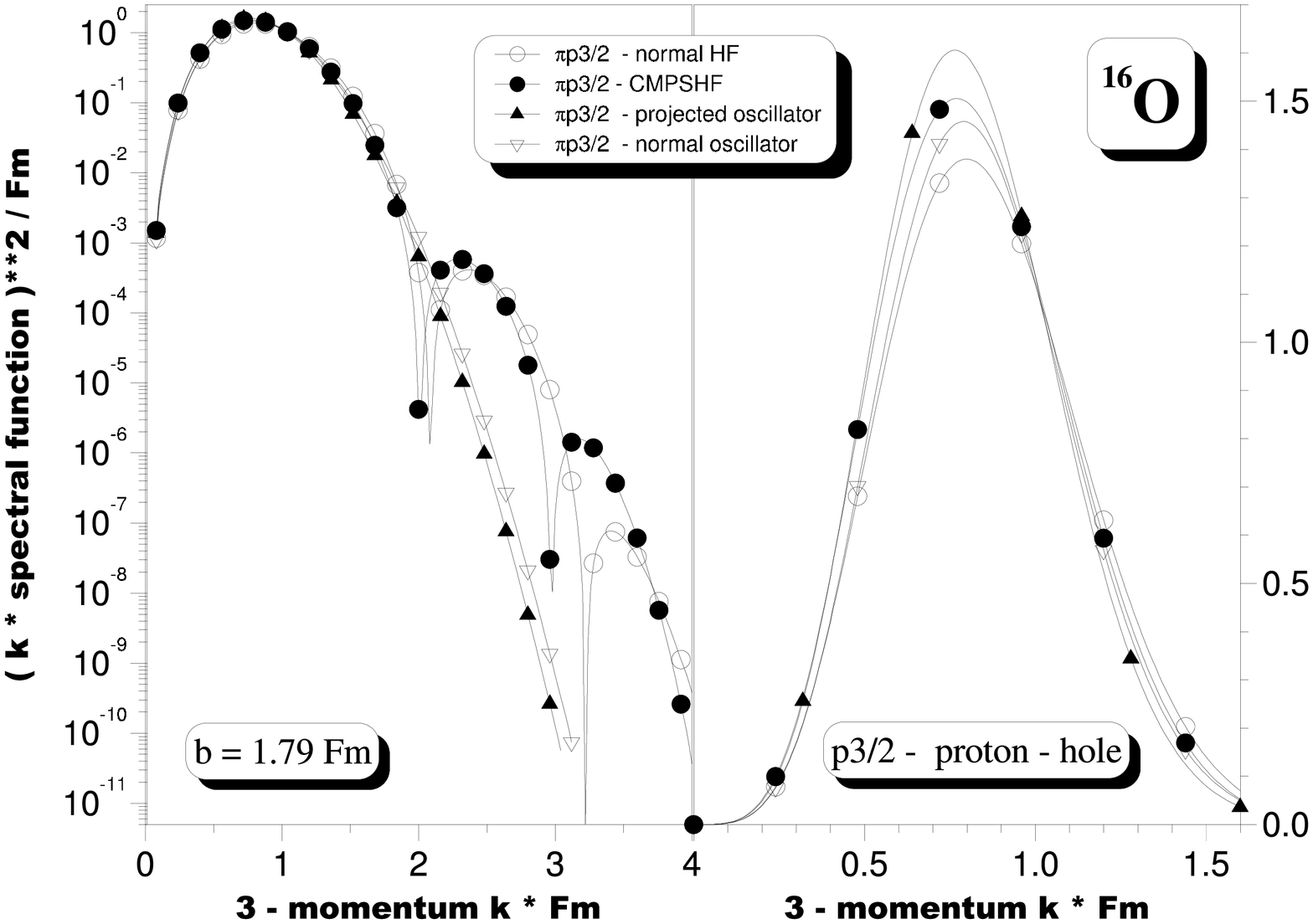}
\caption{
Same as in Fig. 5, but for the p3/2--proton--hole in the nucleus
$^{16}$O.
}
\end{center} 
\end{figure*}

\begin{figure*}
\begin{center}
\includegraphics[angle=-90,width=12cm]{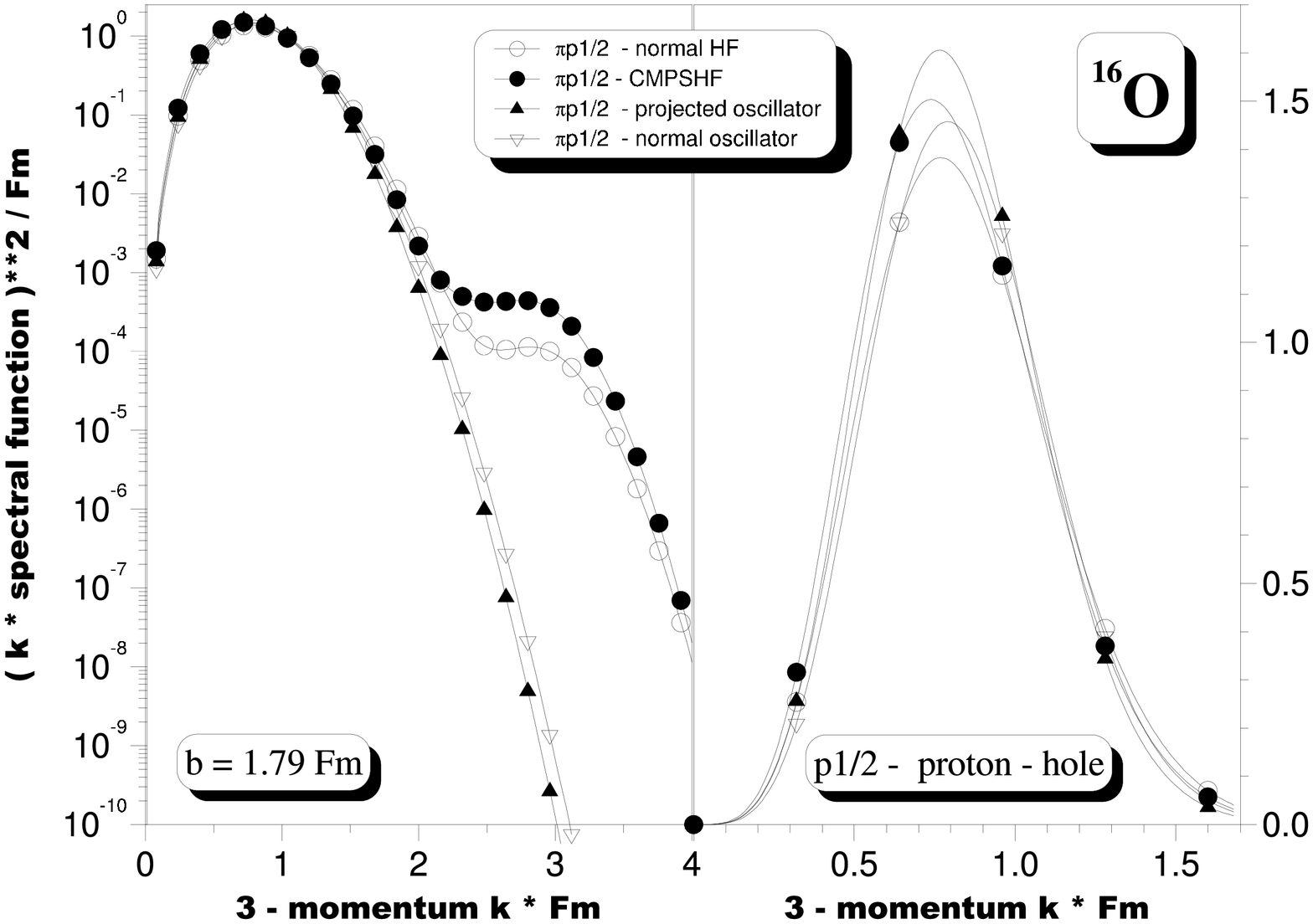}
\caption{
Same as in Fig. 5, but for the p1/2--proton--hole in the nucleus
$^{16}$O.
}
\end{center} 
\end{figure*}

\begin{figure*}
\begin{center}
\includegraphics[angle=-90,width=12cm]{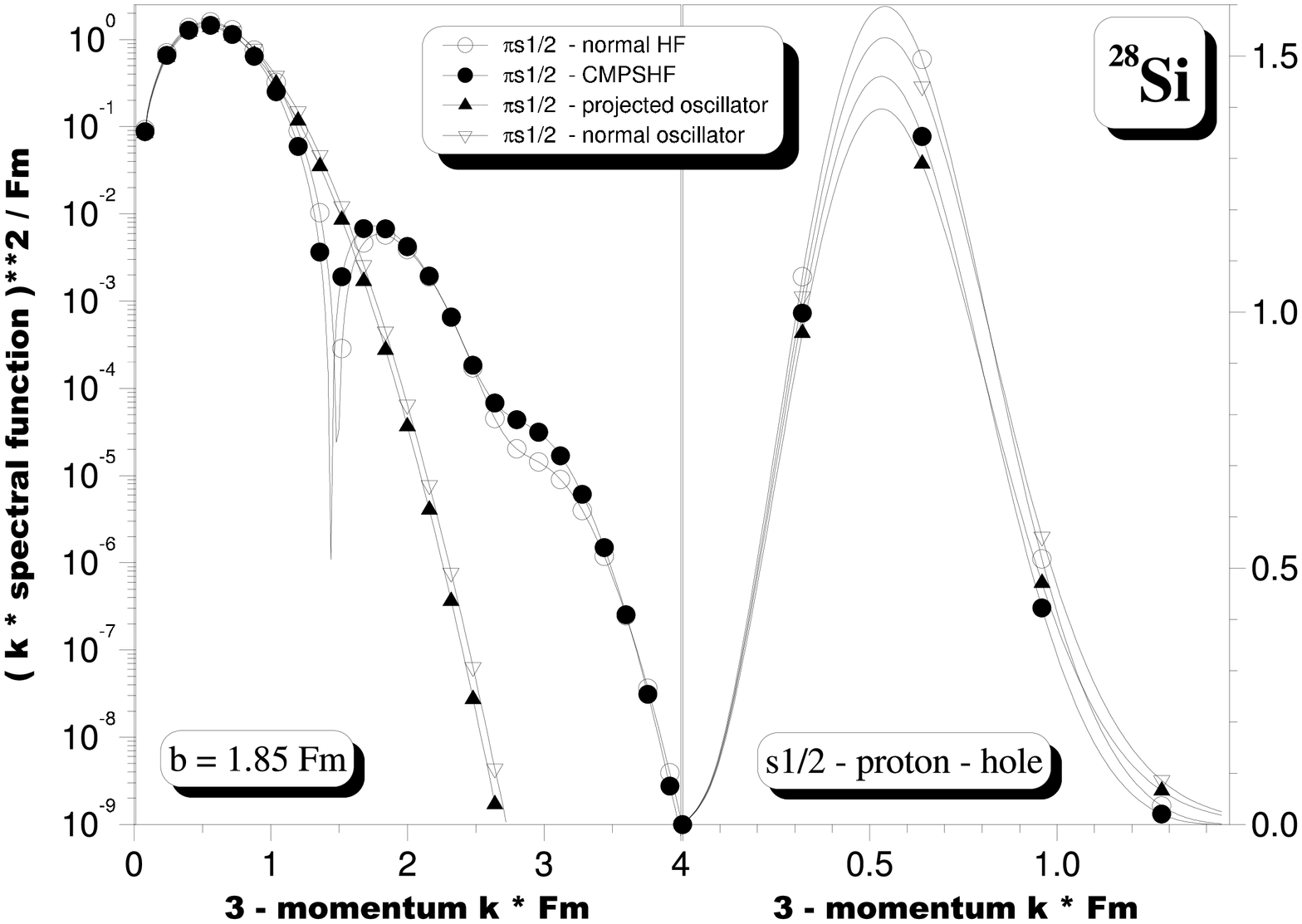}
\caption{
Same as in Fig. 2, but for the s1/2--proton--hole in the nucleus
$^{28}$Si. Here the oscillator length was $b=1.85$ Fm.
}
\end{center} 
\end{figure*}

Obviously, the above description is not Galilei--invariant. First of all,
neither the ground state $\vert D\rangle$ of the considered A--nucleon system
nor the states $b_h\vert D\rangle$ of the (A-1)--nucleon system live in their
respective center of mass rest frame but contain ``spurious'' admixtures
from the motion of the corresponding systems as a whole. In order to obtain a 
Galilei--invariant description, as demonstrated in ref. \cite{ref1.}, instead of
the determinant (\ref{Eq1})
\begin{equation} \label{Eq11}
\vert D;\,0\rangle\,\equiv\,{{\hat C(0)\vert D\rangle}\over
{\sqrt{\langle D\vert\hat C(0)\vert D\rangle}}}
\end{equation}
has to be used as test wave function in the variational calculation yielding
the Hartree--Fock transformation $D$. Here
\begin{equation} \label{Eq12}
\hat C(0)\,\equiv\,\int d^3\vec a\,\hat S(\vec a\,)
\end{equation}
where
\begin{equation} \label{Eq13}
\hat S(\vec a\,)\,\equiv\,\exp\left\{i\vec a\cdot\hat P\right\}
\end{equation}
is the usual shift operator ($\hat P$ being the operator of the total linear
momentum of the considered system), projects $\vert D\rangle$ in its
center of mass rest frame. For the normalization in eq. (\ref{Eq11}) we obtain

\begin{figure*}
\begin{center}
\includegraphics[angle=-90,width=12cm]{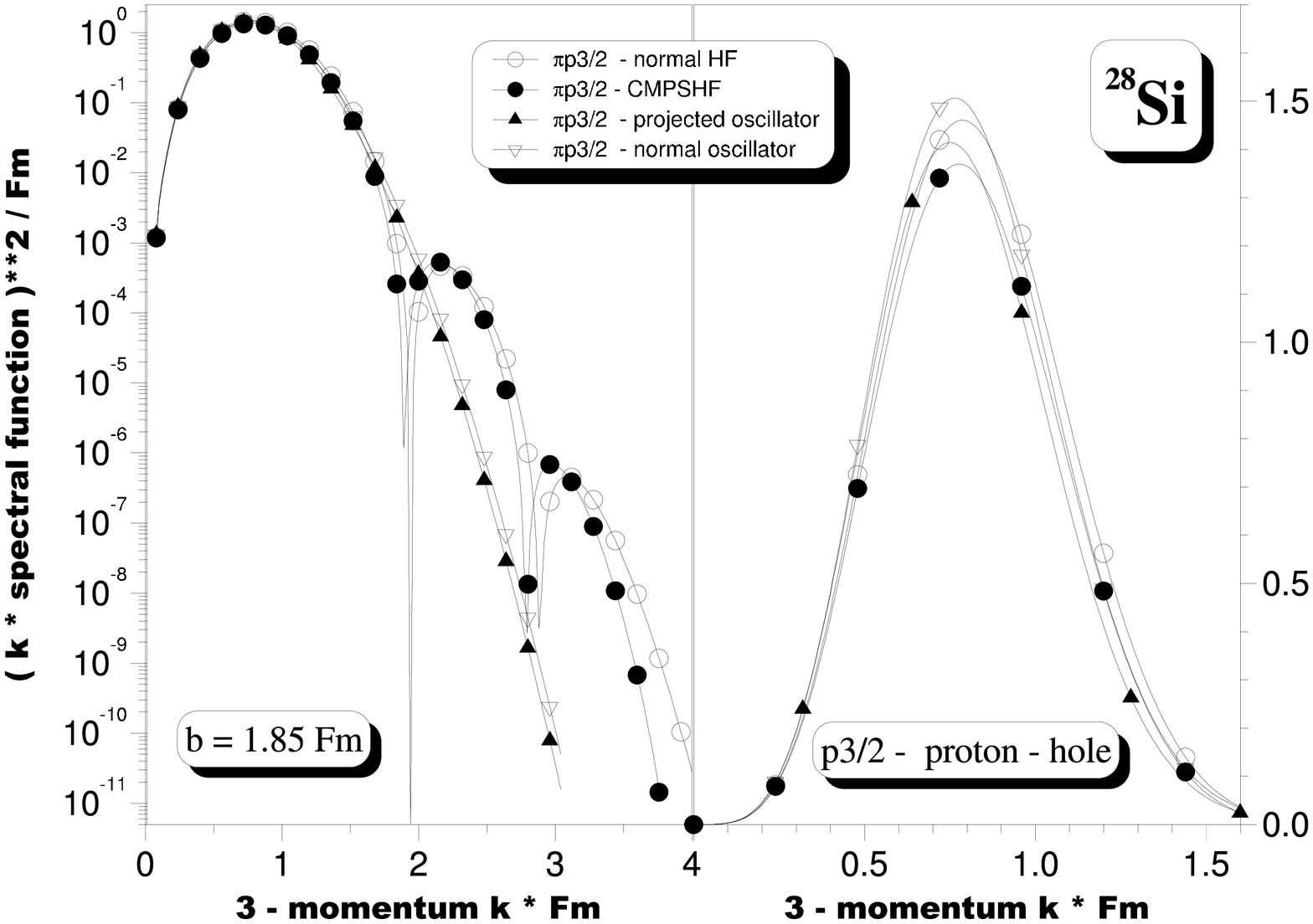}
\caption{
Same as in Fig. 8, but for the p3/2--proton--hole in the nucleus
$^{28}$Si.
}
\end{center} 
\end{figure*}

\begin{figure*}
\begin{center}
\includegraphics[angle=-90,width=12cm]{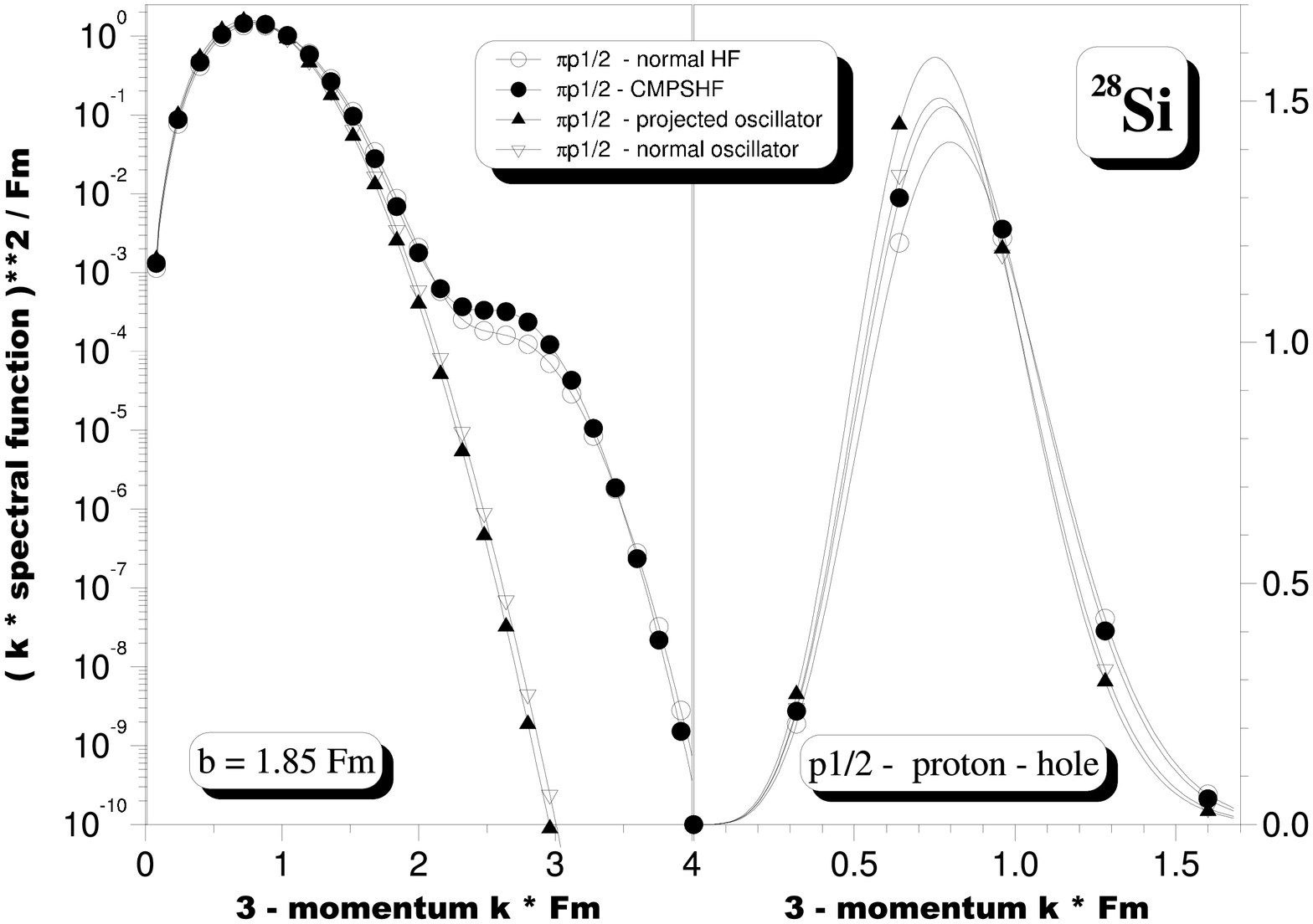}
\caption{
Same as in Fig. 8, but for the p1/2--proton--hole in the nucleus
$^{28}$Si.
}
\end{center} 
\end{figure*}

\begin{eqnarray} \label{Eq14}
N_0 = \langle D\vert\hat C(0)\vert D\rangle
\,=\,\int d^3\vec a\,\langle D\vert\hat S(\vec a\,)\vert D\rangle\cr
=\,4\pi b^3\int\limits_0^{\infty} d\alpha\,\alpha^2\,
\exp\left\{-{A\over 4}\,\alpha^2\right\}\,
{\rm det}\,s(\alpha) 
\nonumber\\
=\,4\pi b^3 \left({4\over A}\right)^{3/2}\,\int\limits_0^{\infty} dy\,
{\rm e}^{-y^2}\,y^2\,
{\rm det}\,s\left({2\over{\sqrt{A}}}y\right)\,
\nonumber\\
=4\pi b^3 \left({4\over A}\right)^{3/2}\,n_0
\end{eqnarray}
where $b$ is the oscillator length parameter, $\vec\alpha\,\equiv\,
\vec a/b$, $y=\alpha\sqrt{A}/2$ and the single particle matrix elements of
the shift operator

\begin{figure*}
\begin{center}
\includegraphics[angle=-90,width=12cm]{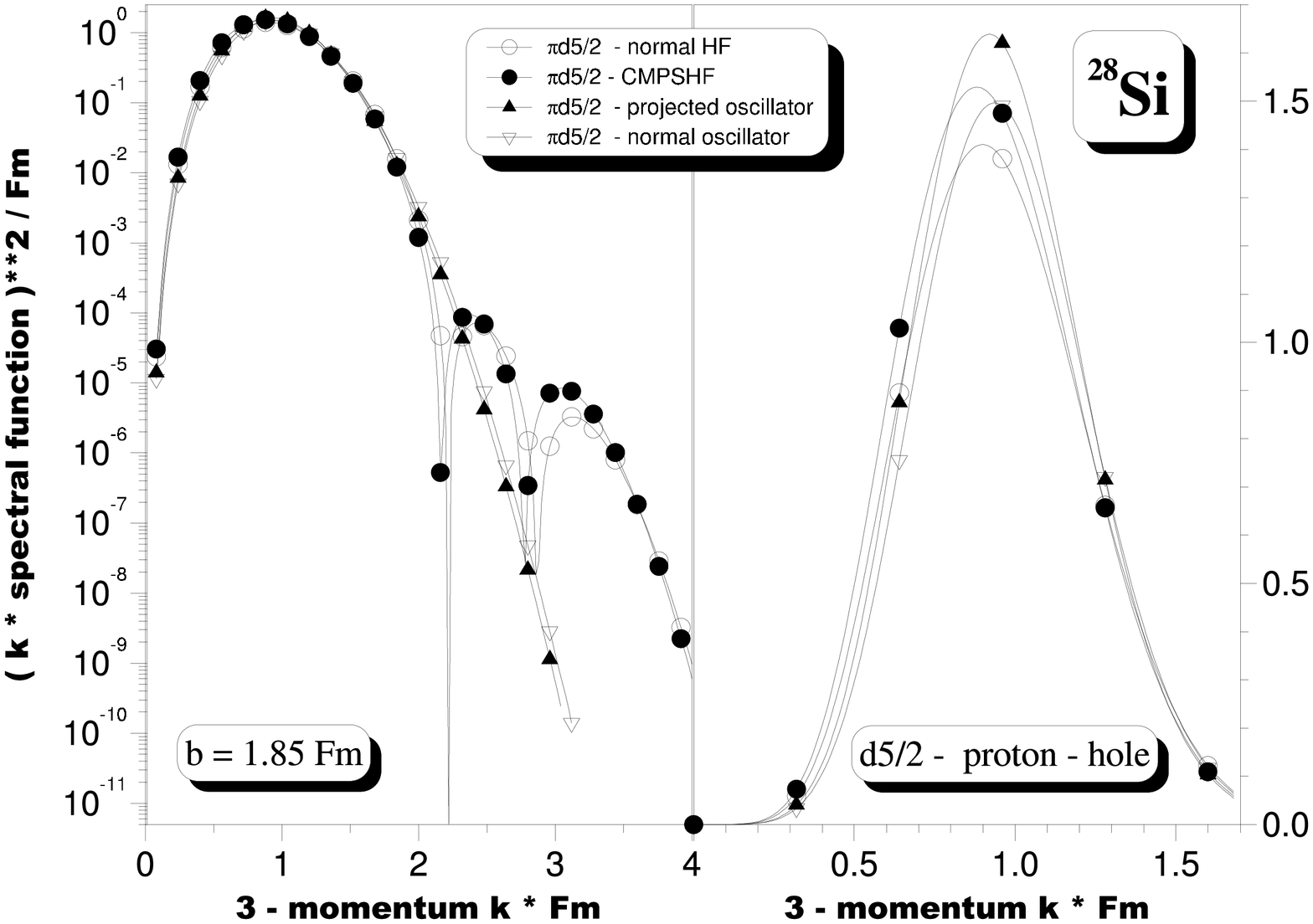}
\caption{
Same as in Fig. 8, but for the d5/2--proton--hole in the nucleus
$^{28}$Si.
}
\end{center} 
\end{figure*}

\begin{figure*}
\begin{center}
\includegraphics[angle=-90,width=12cm]{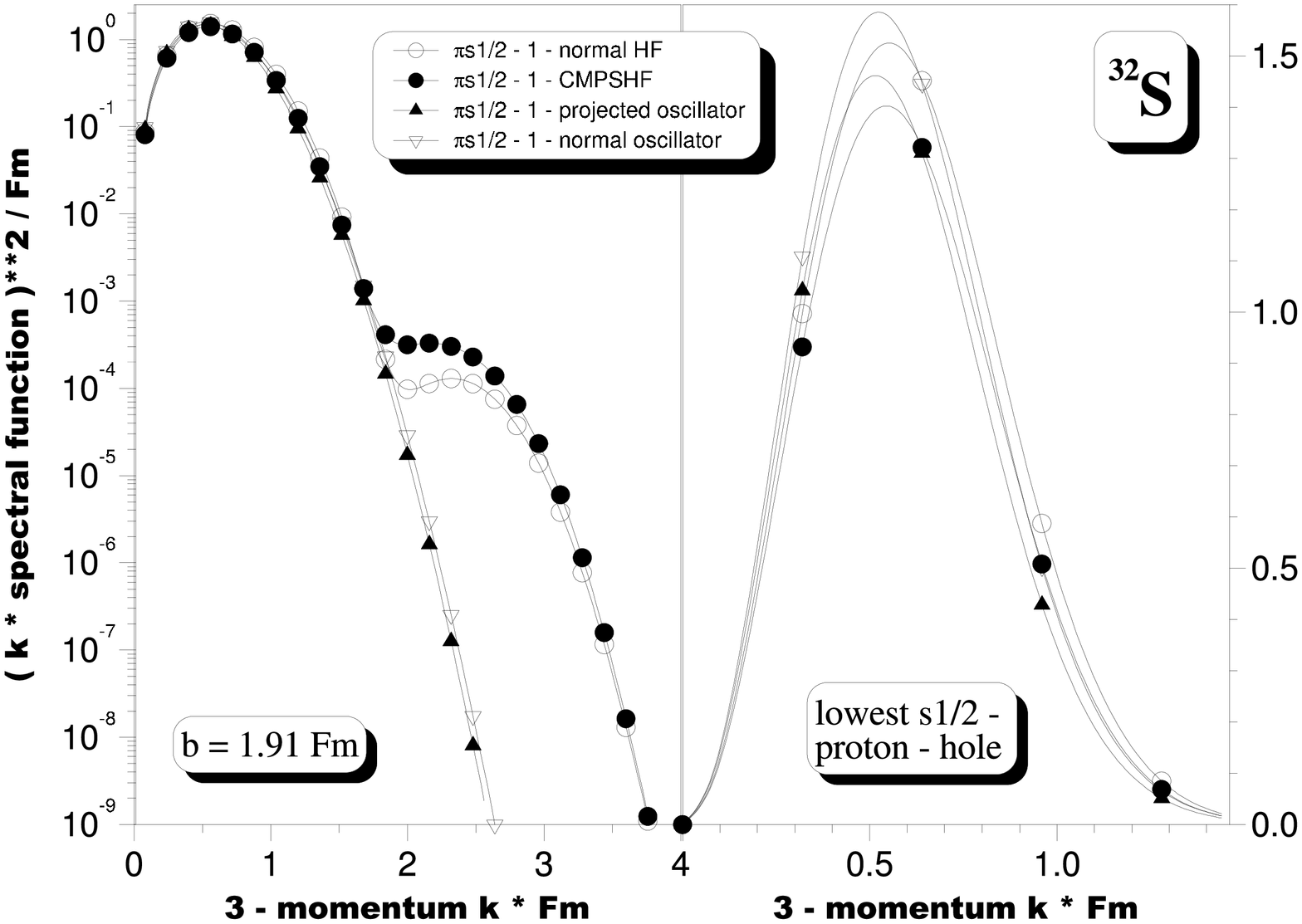}
\caption{
Same as in Fig. 2, but for the lowest s1/2--proton--hole in the
nucleus $^{32}$S. Here the oscillator length was $b=1.91$ Fm.
}
\end{center} 
\end{figure*}

\begin{eqnarray} \label{Eq15}
s_{hh^{\prime}}(\alpha)\,\equiv\,{\rm e}^{\alpha^2/4}\,
\langle h\vert\hat S(\hat e_z\cdot\vec a\,)\vert h^{\prime}\rangle
\end{eqnarray}
have been given in ref. \cite{ref1.}. In eqs. (\ref{Eq14}) and (\ref{Eq15}) we have used that
$\vert D\rangle$ is spherically symmetric (i.e., invariant under rotations) so
that we can put the shift vector without loss of generality into the
z--direction.

\begin{figure*}
\begin{center}
\includegraphics[angle=-90,width=12cm]{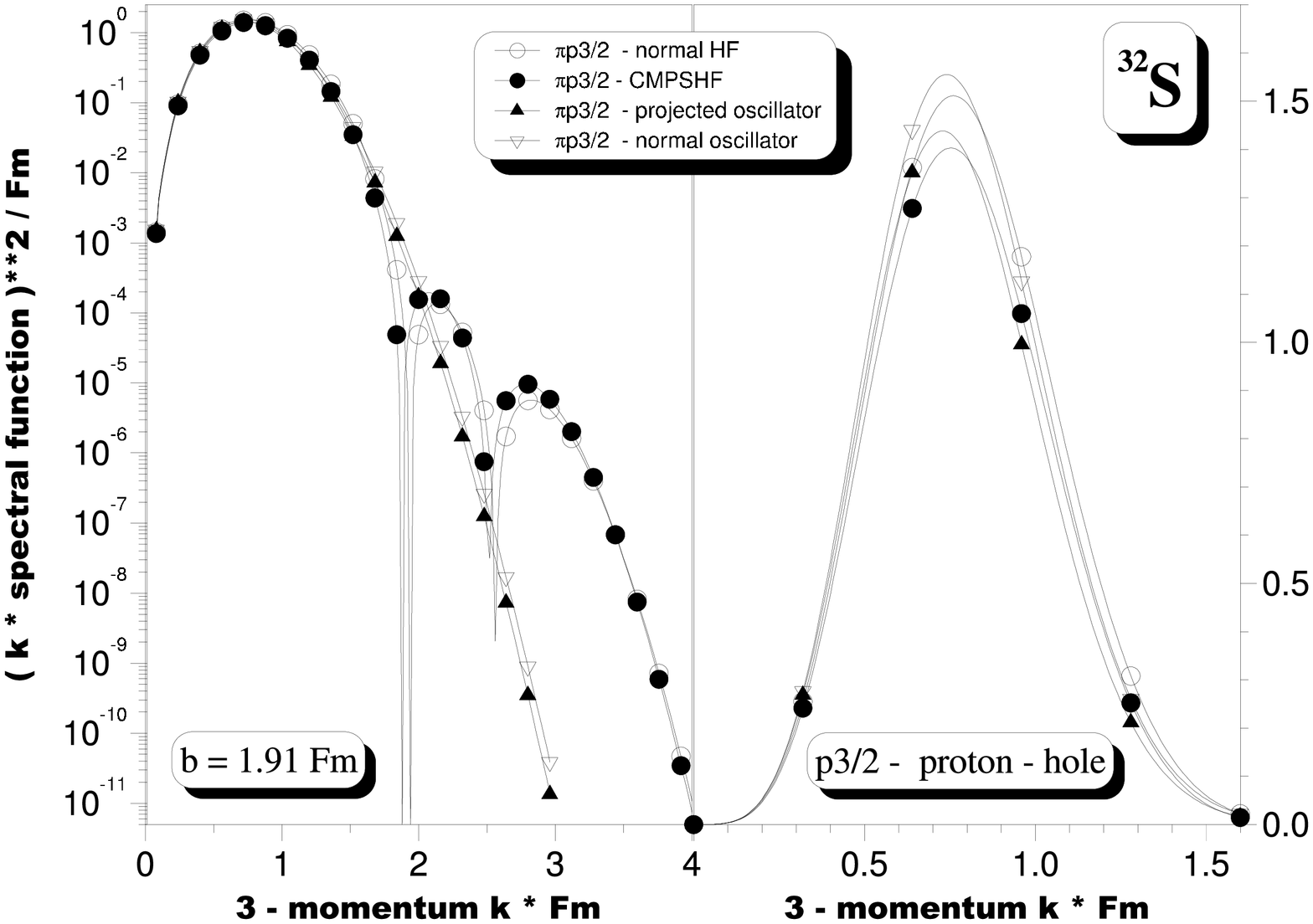}
\caption{
ame as in Fig. 12, but for the p3/2--proton--hole in the nucleus
$^{32}$S.
}
\end{center} 
\end{figure*}

\begin{figure*}
\begin{center}
\includegraphics[angle=-90,width=12cm]{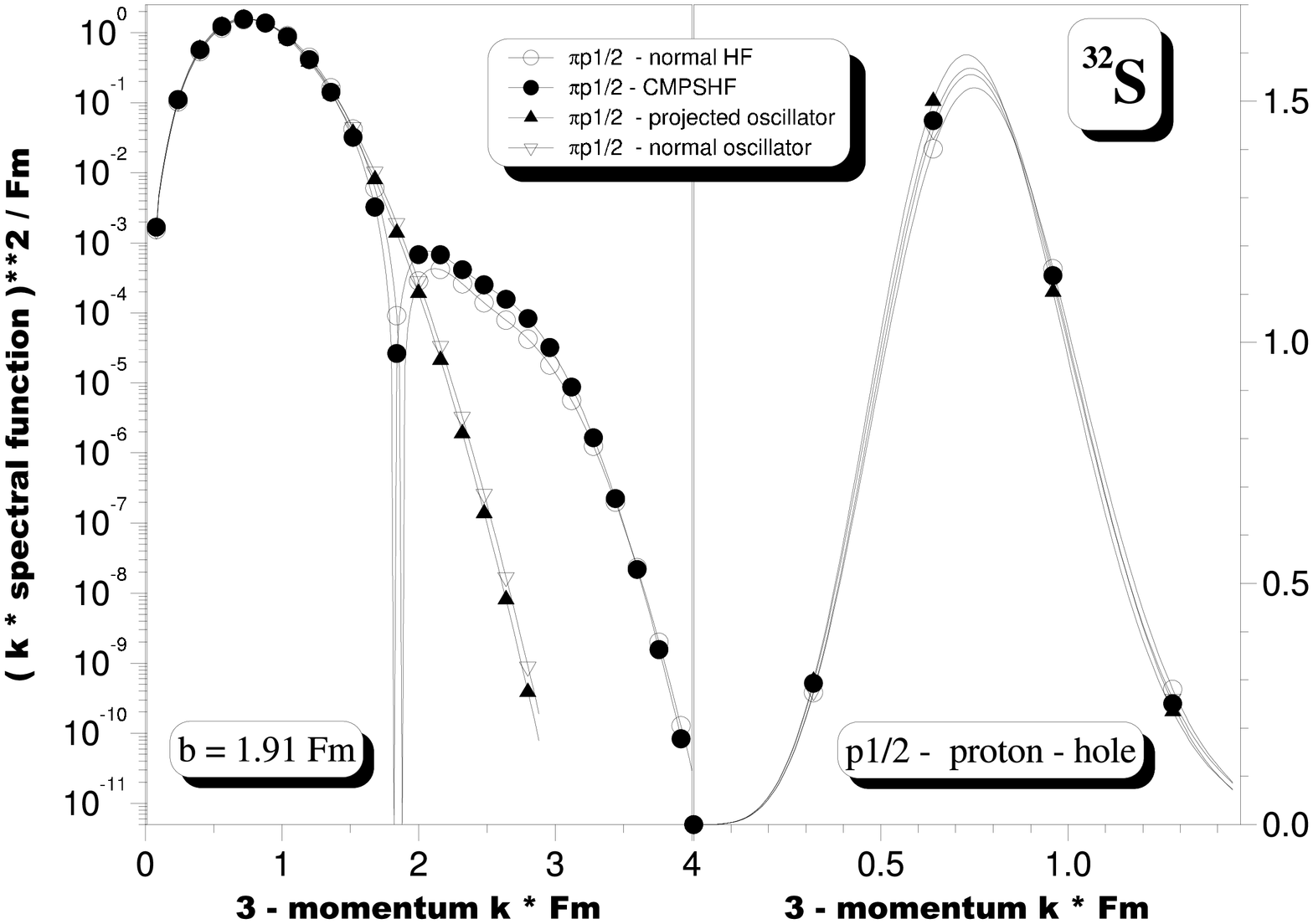}
\caption{
Same as in Fig. 12, but for the p1/2--proton--hole in the nucleus
$^{32}$S.
}
\end{center} 
\end{figure*}

In the same way we define

\begin{eqnarray} \label{Eq16}
N_{hh^{\prime}}\,\equiv\,\langle D\vert b^{\dagger}_h\hat C(0) b_{h^{\prime}}
\vert D\rangle\,\equiv\,4\pi b^3 \left({4\over{A-1}}\right)^{3/2}\,
n_{hh^{\prime}}
\nonumber\\
\end{eqnarray} 
and

\begin{eqnarray} \label{Eq17}
H_{hh^{\prime}}\,\equiv\,\langle D\vert b^{\dagger}_h\hat H
\hat C(0) b_{h^{\prime}}
\vert D\rangle\,\equiv\,4\pi b^3 \left({4\over{A-1}}\right)^{3/2}\,
h_{hh^{\prime}}
\nonumber\\
\end{eqnarray}
where $\hat H$ denotes the same Hamiltonian as has been used to obtain the
Hartree--Fock transformation $D$. Solving the generalized eigenvalue problem
\begin{eqnarray} \label{Eq18}
(h-En)w\,=\,0\qquad{\rm with}\qquad w^Tnw\,\equiv\,{\bf 1}
\end{eqnarray}
the Galilei--invariant form of the one--hole states can be written as
\begin{eqnarray} \label{Eq19}
\vert\tilde h;\,0\rangle\,=\,{\sum\limits_{\alpha_h}}^{(\tau_h l_h j_h)}
\hat C(0) b_h\vert D\rangle\,w^{\tau_h l_h j_h}_{\alpha_h\tilde h}\,
{1\over{\sqrt{4\pi b^3 (4/(A-1))^{3/2}}}}
\nonumber\\
\end{eqnarray}

Furthermore, for the outgoing (or incoming) continuum nucleon not the state (\ref{Eq4})
but a {\it relative} wave function with respect to the (A-1)--nucleon system
should be used. Consequently, we get for the Galilei--invariant hole--spectral
function

\begin{eqnarray} \label{Eq20}
f_{\tilde h\tau\sigma}^{proj}(\vec k\,) =
\delta_{\tau\tau_h}\,{{n_0^{-1/2}}\over{4\pi b^3(4/A)^{3/4}(4/(A-1))^{3/4}}}
\times
\nonumber\\
\times
{\sum\limits_{\alpha_h}}^{(\tau_h l_h j_h)}
\langle D\vert C^{\dagger}_{\vec k\,\tau\sigma}
\exp\{-i\vec k\cdot\vec R_{A-1}\}
\hat C(0) b_h\vert D\rangle\
w^{\tau_h l_h j_h}_{\alpha_h\tilde h}
\nonumber\\
=\delta_{\tau\tau_h}\,{{n_0^{-1/2}}\over{4\pi b^3(4/A)^{3/4}(4/(A-1))^{3/4}}}
{\sum\limits_{\alpha_h}}^{(\tau_h l_h j_h)}
{\sum\limits_{\alpha_{h^{\prime}} l_{h^{\prime}} j_{h^{\prime}}
m_{h^{\prime}}}}^{(\tau_h)}
\nonumber\\
\langle h^{\prime}\vert\vec k\,\tau_h\sigma\rangle
\langle D\vert b^{\dagger}_{h^{\prime}}
\exp\{-i\vec k\cdot\vec R_{A-1}\}
\hat C(0) b_h\vert D\rangle\,w^{\tau_h l_h j_h}_{\alpha_h\tilde h}
\nonumber\\
\end{eqnarray}
where $\vec R_{A-1}$ denotes the center of mass coordinate of the
(A-1)--nucleon system. Again the spherical symmetry of $\vert D\rangle$
and the properties of the hole--creation and --annihilation operators
under rotations can be used to put for the evaluation of the matrix element
in eq. (\ref{Eq20}) one vector without loss of generality in the z--direction. Here
we choose $\vec k\,=\,\hat e_z\cdot\vec k$. In analogy to eq. (\ref{Eq6}) we can write
\begin{eqnarray} \label{Eq21}
f_{\tilde h\tau\sigma}^{proj}(\vec k\,)\,=\,\delta_{\tau\tau_h}\,
i^{l_h}\,\sum\limits_{\lambda_h}\,Y_{l_h\lambda_h}^*(\hat k)
(l_h 1/2 j_h\vert \lambda_h\sigma m_h)\,g_{\tau_h\tilde h\,l_h j_h}^{proj}(k)
\nonumber\\
\end{eqnarray}
where the ``reduced'' projected spectral function is now given by

\begin{eqnarray} \label{Eq22}
g_{\tau_h\tilde h\,l_h j_h}^{proj}(k) = \left({A\over{A-1}}\,b^2\right)^{3/4} 
 \exp\left\{-{1\over 2}{A\over{A-1}}(bk)^2\right\}  n_0^{-1/2}
 \nonumber\\ 
{\sum\limits_{\alpha_h}}^{(\tau_h l_h j_h)} 
w^{\tau_h l_h j_h}_{\alpha_h\tilde h}
{\sum\limits_{l_{h^{\prime}} j_{h^{\prime}}\alpha_{h^{\prime}}}}^{(\tau_h)}
\tilde R^{\tau_h l_{h^{\prime}} j_{h^{\prime}}}_{\alpha_{h^{\prime}}}(bk)
\nonumber\\
\sum\limits_L \Delta(l_h,\,l_{h^{\prime}},\,L) {1\over 2}\left[
1+(-)^{l_h+l_{h^{\prime}}+L}\right] \sqrt{{{2j_{h^{\prime}}+1}\over{2j_h+1}}}
\nonumber\\
(-)^{j_{h^{\prime}}-1/2}(j_{h^{\prime}} j_h L\vert 1/2 -1/2 0) 2 
\nonumber\\
\sum\limits_{m>0}(-)^{j_h-m}(j_{h^{\prime}} j_h L\vert m -m 0)
\nonumber\\
exp\left\{{1\over{A-1}}\left({{bk}\over 2}\right)^2\right\}
\int\limits_0^{\infty} dy {\rm e}^{-y^2} y^2\int\limits_0^{\pi/2}
d\vartheta\sin\vartheta
\nonumber\\
\Biggl\{{1\over 2}[1+(-)^{l_h+l_{h^{\prime}}}](-)^{(l_{h^{\prime}}
-l_h)/2} \times
\nonumber\\
{\rm Re}\Biggl[\left(z^{(\tau_h)}(bk,\alpha,\vartheta)
\right)^{-1}_{\alpha_h l_h j_h m;\,
\alpha_{h^{\prime}} l_{h^{\prime}} j_{h^{\prime}}m}
\,{\rm det}\,z\Biggr] +
\nonumber\\
+\,{1\over 2}[1-(-)^{l_h+l_{h^{\prime}}}](-)^{(l_{h^{\prime}}
-l_h-1)/2} \times
\nonumber\\
{\rm Im}\Biggl[\left(z^{(\tau_h)}(bk,\alpha,\vartheta)
\right)^{-1}_
{\alpha_h l_h j_h m;\,\alpha_{h^{\prime}} l_{h^{\prime}} j_{h^{\prime}}m}
\,{\rm det}\,z\Biggr]\Biggr\}
\nonumber\\
\end{eqnarray} 
Here $\Delta(l_h,\,l_{h^{\prime}},\,L)=1$ if $\vert l_h-l_{h^{\prime}}\vert
\leq L\leq l_h+l_{h^{\prime}}$ and $=0$ if not, $y\,\equiv\,
\alpha\sqrt{A-1}/2$, $\vartheta$ is the angle between
the shift vector and the z--direction and

\begin{eqnarray}\label{Eq23}
\tilde R^{\tau_h l_{h^{\prime}} j_{h^{\prime}}}_{\alpha_{h^{\prime}}}(bk)
\,\equiv\,\sum\limits_{n^{\prime}}\,(-)^{n^{\prime}}\,
D^{\tau_h l_{h^{\prime}} j_{h^{\prime}}}_{n^{\prime}\alpha_{h^{\prime}}}\,
\tilde R_{n^{\prime} l_{h^{\prime}}}(bk)
\noindent
\end{eqnarray}
denotes the polynomial part of the Fourier--transform of the radial
wave function of the hole state $h^{\prime}$. The matrix elements of 
$z^{(\tau_h)}$ are identical to those given in appendix B of ref. \cite{ref1.}
except for the fact that the argument $qb/A$ has to be replaced here by
$kb/(A-1)$. The determinant factorizes in a proton-- and a neutron--part

\begin{equation} \label{Eq24}
{\rm det}z\,=\,\prod\limits_{\tau=p,n}\,{\rm det}z^{(\tau)}
\end{equation}
and obviously depends on the same arguments as the matrix elements of the
inverse matrices $(z^{(\tau)})^{-1}$ in eq. (\ref{Eq22}).

\begin{figure*}
\begin{center}
\includegraphics[angle=-90,width=12cm]{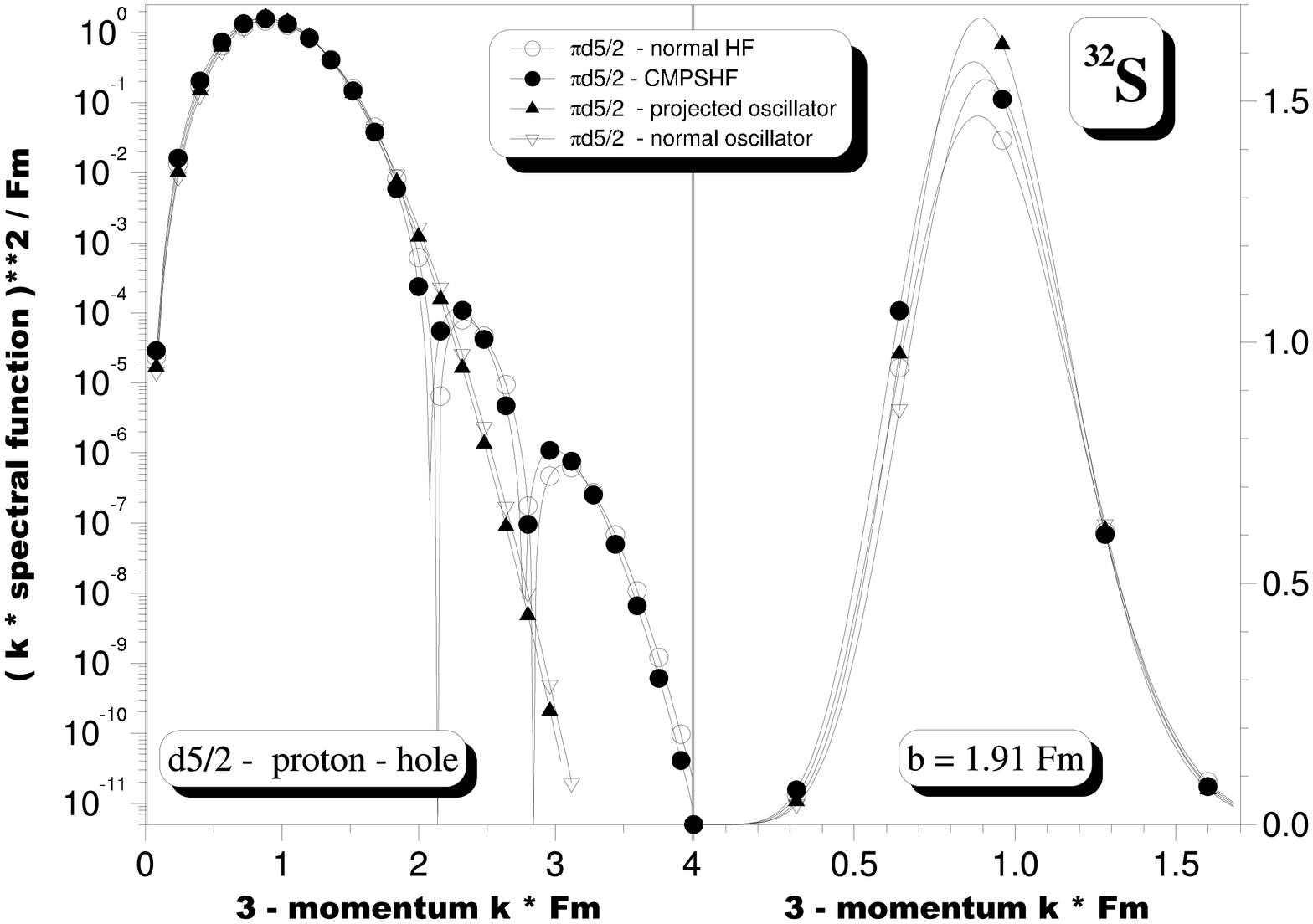}
\caption{
Same as in Fig. 12, but for the d5/2--proton--hole in the nucleus
$^{32}$S.
}
\end{center} 
\end{figure*}

Similarly as in eq. (\ref{Eq9}), the projected spectroscopic factors are then defined
by

\begin{eqnarray} \label{Eq25}
S_{\tilde h}^{proj}\,\equiv\,\sum\limits_{\sigma}\int d^3\vec k\,
\vert f_{\tilde h\tau\sigma}^{proj}(\vec k\,)
\vert^2\,
\nonumber\\
=\delta_{\tau\tau_h}\int\limits_0^{\infty} dk\,k^2\,
{g_{\tau_h\tilde h\,l_h j_h}^{proj}(k)\,}^2
\end{eqnarray}

It has been demonstrated in ref. \cite{ref2.} that, if the hole states
 (\ref{Eq2}) are pure
harmonic oscillator states and if $\vert D\rangle$ is a ``non--spurious''
oscillator determinant, then eqs. (\ref{Eq22}) and (\ref{Eq25}) can be 
evaluated analytically.
In this case one does not even have to solve the generalized eigenvalue
problem (\ref{Eq18}). In all doubly closed j--shell nuclei up
 to $^{28}$Si, the overlap
matrix $n_{hh^{\prime}}$ out of eq. (\ref{Eq16}) is diagonal, so that the $w$'s
needed in eq. (\ref{Eq21}) are simply the inverse square roots of its diagonal
elements. In $^{32}$S and $^{40}$Ca two s--states are occupied and do mix
via eq. (\ref{Eq16}). Here one can always Gram--Schmidt--orthonormalize the 
0s--state with respect to the 1s--state as has been shown in ref. \cite{ref2.}.
The result of these analytical calculations was that the (reduced) hole
spectral function (\ref{Eq22}) can be written as

\begin{equation} \label{Eq26}
g_{\tau_h\tilde h\,l_h j_h}^{proj;\,osc}(k)\,\equiv\,
R^{rel}_{\tilde h l_h}(k)\,\sqrt{S^{proj;\,osc}_{\tilde h}}
\end{equation}
where, in the one--dimensional cases, the subscript $\tilde h\,=\,n_h$ and

\begin{eqnarray} \label{Eq27}
R^{rel}_{n_h l_h}(k) = (-)^{n_h}
\left({A\over{A-1}}\,b^2\right)^{3/4}
\nonumber\\
\exp\left\{-{1\over 2}{A\over{A-1}}
(bk)^2\right\}\,\tilde R_{n_h l_h}\left(\sqrt{{A\over{A-1}}}bk\right),
\end{eqnarray}
while for the lowest s--states in $^{32}$S and $^{40}$Ca the 
$R^{rel}_{\tilde h l_h}$ are linear combinations of the functions (\ref{Eq27})
for $l_h=0$ and $n_h=0,\,1$ with the corresponding expansion coefficients
resulting from the orthonormalization. The functions (\ref{Eq27}) are just
the usual oscillator wave functions in momentum representation, however,
the nucleon mass $M$ entering the parameter $b$ has been replaced by the
reduced mass $(A-1)M/A$. This is indicated by the superscript
$rel$. The analytical evaluation in ref. \cite{ref2.} yielded

\begin{eqnarray}\label{Eq28}
S^{proj;\,osc}_{\tilde h}\,=\,\left\{\matrix{
1 &{\rm for}\;0s-{\rm holes}\;{\rm in}\;^4{\rm He}\cr
 &\cr
{4\over 5} &{\rm for}\;0s-{\rm holes}\;{\rm in}\;^{16}{\rm O}\cr
 &\cr
{{16}\over{15}} &{\rm for}\;0p-{\rm holes}\;{\rm in}\;^{16}{\rm O}\cr
 &\cr
{{1410}\over{1521}} &{\rm for}\;0\tilde s-{\rm holes}\;{\rm in}\;^{40}{\rm Ca}
\cr
 &\cr
{{1400}\over{1521}} &{\rm for}\;0p-{\rm holes}\;{\rm in}\;^{40}{\rm Ca}\cr
 &\cr
{{1600}\over{1521}} &{\rm for}\;1s-{\rm holes}\;{\rm in}\;^{40}{\rm Ca}\cr
 &\cr
{{1600}\over{1521}} &{\rm for}\;0d-{\rm holes}\;{\rm in}\;^{40}{\rm Ca}\cr}
\right\}
\end{eqnarray}
It is easily checked that
\begin{eqnarray}\label{Eq29}
\sum\limits_{\tilde h}\,S^{proj;\,osc}_{\tilde h}\,=\,A
\end{eqnarray}
as expected, since the functions (27) form again a complete orthonormal
set. On first sight it may seem strange to obtain spectroscopic factors
which are larger than one, however, the oscillator results (28) are
identical to those obtained by Dieperink and de Forest \cite{ref10.} with rather
different methods and are easy to understand by simple considerations
as discussed in ref. \cite{ref2.}.

\begin{figure*}
\begin{center}
\includegraphics[angle=-90,width=12cm]{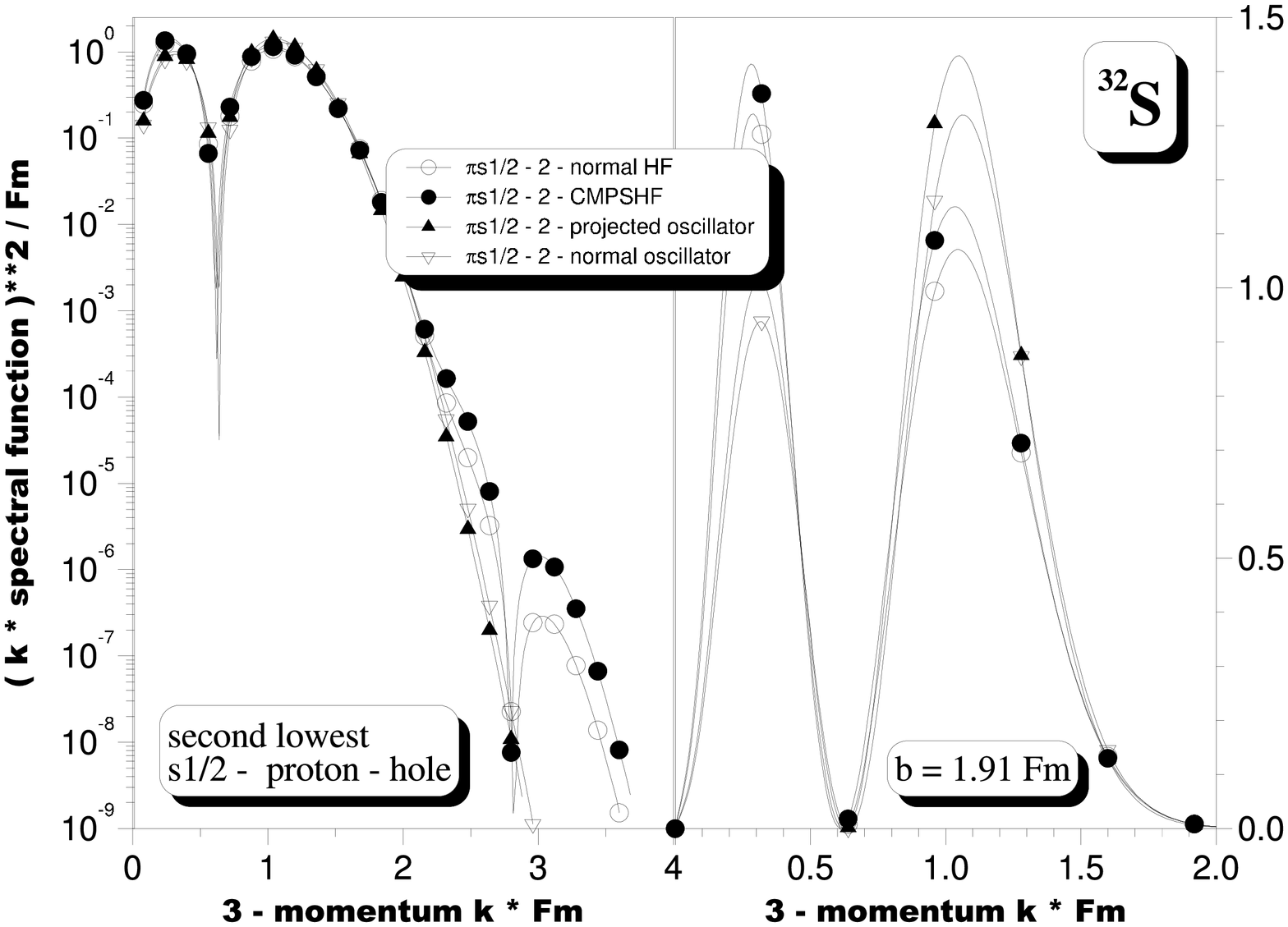}
\caption{
Same as in Fig. 12, but for the second lowest s1/2--proton--hole
in the nucleus $^{32}$S.
}
\end{center} 
\end{figure*}

Unlike the ``normal'' result (\ref{Eq7}) and the projected oscillator 
result (\ref{Eq26})
for non--spurious oscillator determinants $\vert D\rangle$, the general
form of the reduced spectral function (\ref{Eq22}) cannot be written as the
product of a normalized single particle wave function times some number. Thus
the sum of the spectroscopic factors over all hole--states does in
general yield a smaller number than A, since
$\exp\{-i\vec k\cdot\vec R_{A-1}\}C_{\vec k\,\tau\sigma}\vert D\rangle$
does not contain the complete set of configurations created by
$\hat C(0)\,b_h\vert D\rangle$. We therefore write

\begin{equation} \label{Eq30}
\sum\limits_{\tilde h}\,S^{proj}_{\tilde h}\,=\,A\,-\,\epsilon
\end{equation}
It turns out, however, that for the cases discussed in the next section
$\epsilon/A$ varies only between 0.12 and 0.35 percent. The violation
of the sum rule due to non--trivial correlations induced by the projector into
the uncorrelated Hartree--Fock systems investigated here is hence rather
small, and at least approximately, the separation of eq. (\ref{Eq22}) into a product
of a normalized single particle function and the square root of the
spectroscopic factor is still true.

\section{Results and discussion.}
We considered here the six doubly closed j--shell nuclei $^4$He, $^{12}$C,
$^{16}$O, $^{28}$Si, $^{32}$S and $^{40}$Ca. In ref. \cite{ref1.} for these nuclei
the results of Galilei--invariant spherical Hartree--Fock calculations with
projection into the center of mass rest frame have been compared to those
of standard spherical Hartree--Fock calculations, in which only the kinetic
energy of the center of mass motion was subtracted from the hamiltonian
before the variation. As hamiltonian the kinetic energy plus the Coulomb
interaction plus the Brink--Boeker force B1 [6] complemented with a short
range (0.5 Fm) two--body spin--orbit term having the same volume integral
as the Gogny--force D1S \cite{ref7.} has been used. The Hartree--Fock wave functions
resulting from these calculations using as single particle basis 19 major
oscillator shells will be analyzed in the following. Furthermore, as in
ref. \cite{ref1.}, also here the results will be compared to those obtained with
simple oscillator determinants.

We shall first discuss the hole--spectroscopic factors. Since in the normal
approach these are all equal to one (see eq. (\ref{Eq9})), irrespective whether 
one uses simple oscillator determinants or the Hartree--Fock ground states
$\vert D_c\rangle$ out of eq. (\ref{Eq27}) in ref. \cite{ref1.}, we can restrict
 ourselves to
the discussion of the Galilei--invariant spectroscopic factors out of
eqs. (\ref{Eq26}) and (\ref{Eq25}) for the oscillator occupations and 
for the Hartree--Fock
ground states $\vert D_{pr}\rangle$ (see eq. (\ref{Eq28}) of ref. \cite{ref1.}) obtained 
with
projection into the center of mass rest frame before the variation,
respectively. Since furthermore in the oscillator approach proton-- and
neutron--hole--spectroscopic factors are identical and even in the
Hartree--Fock approach almost undistiguishable, only the proton--hole
spectroscopic factors will be discussed.

\begin{figure*}
\begin{center}
\includegraphics[angle=-90,width=12cm]{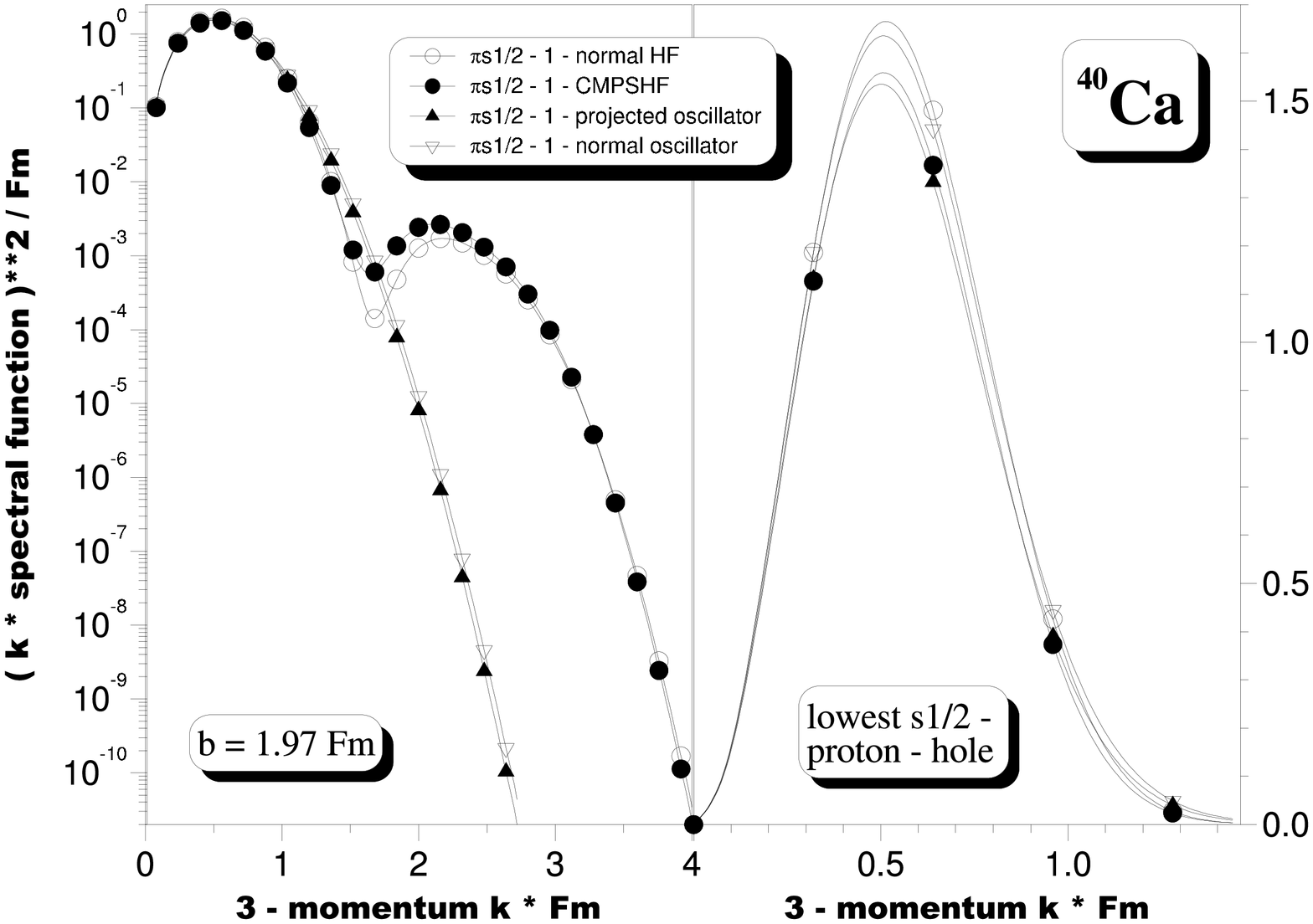}
\caption{
Same as in Fig. 2, but for the lowest s1/2--proton--hole in the
nucleus $^{40}$Ca. Here the oscillator length was $b=1.97$ Fm.
}
\end{center} 
\end{figure*}

\begin{figure*}
\begin{center}
\includegraphics[angle=-90,width=12cm]{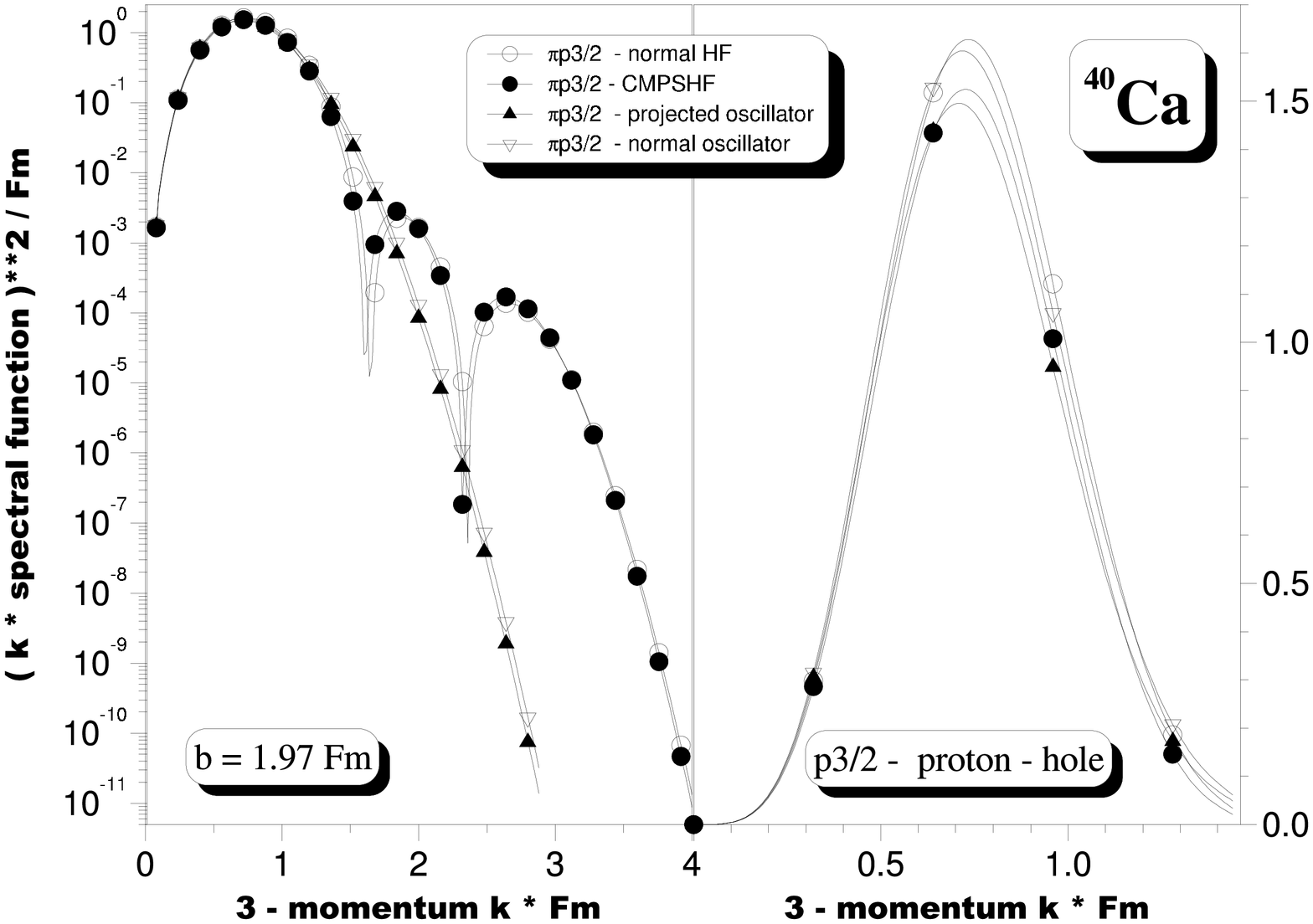}
\caption{
Same as in Fig. 17, but for the p3/2--proton--hole in the nucleus
$^{40}$Ca.
}
\end{center} 
\end{figure*}

The results are summarized in figure 1. Open symbols refer
to the projected oscillator, closed (or crossed) symbols to the projected
Hartree--Fock results. On the abscissa the relevant orbits are presented.
They are pure oscillator orbits (or for the 0s1/2--states in $^{32}$S and
$^{40}$Ca the states resulting from Gram--Schmidt orthonormalization
with respect to the 1s1/2--states) in the former, the lowest (``0'') or
second lowest (``1'') Hartree--Fock single particle states resulting from
the solution of eq. (18) for each set of $l$ and $j$ quantum numbers in the
latter case. The figure clearly shows that, though based on rather different
wave functions, oscillator and Hartree--Fock results are almost identical.
Thus for the spectroscopic factors, which are integral properties, the choice
of the single particle basis, which is irrelevant in the normal approach, does
not seem to matter in the Galilei--invariant prescription either, at least as
long as only uncorrelated systems are considered. As already discussed in
ref. \cite{ref2.}, one sees a considerable depletion of the strengths of the
hole--states with excitation energies larger or equal to $1\hbar\omega$ and an
enhancement of the strengths of the hole--orbits near the Fermi--energy. The
depletion of the lowest s--state in $^{16}$O is as large as 20 percent and
even for the lowest s-- and p3/2--states in $^{32}$S and $^{40}$Ca still
depletions of more than 8 percent are obtained. Exceptions of the
$\hbar\omega$--rule are the
spectroscopic factors of the p1/2--holes in $^{28}$Si and $^{32}$S. Though
belonging to the second but last shell below the Fermi--energy, these are
considerably less affected by the projection than their p3/2--spin--orbit
partners. A similar spin--orbit effect for these orbits was also seen in the
corresponding single particle energies discussed in ref. [1]. It results from
the presence of the d5/2 and the absence of the d3/2--orbit in both these
nuclei and indicates the dominance of couplings to angular momentum one
in the spurious center of mass motion.

As already discussed, the oscillator results fulfill the sum rule (29)
exactly, while for the Hartree--Fock this is only approximately true
(see eq. (\ref{Eq30})). However, from the fact that oscillator and Hartree--Fock
spectroscopic factors are almost identical, it is clear that the violation
$\epsilon$ of the sum rule (\ref{Eq30}) is, as mentioned before, only a rather small
effect. It was furthermore discussed in ref. \cite{ref4.} that in the oscillator
approach both the normal spectroscopic factors (all equal to one) together
with the normal single particle energies as well as the projected
spectroscopic factors together with the projected single particle
energies fulfill Kolthun's sum rule \cite{ref11.} exactly, which was interpreted
as a nice check for the consistency of the projected results. Now, in case of
the harmonic oscillator approach, the projected total energy for the
A--nucleon ground state is identical to that of the normal approach,
provided the latter is corrected by subtracting the kinetic energy of the
center of mass motion. As has been demonstrated in ref. \cite{ref1.}, in the
Hartree--Fock prescription this is not the case. Thus, while in the normal
Hartree--Fock approach Kolthun's sum rule is fulfilled by definition, in the
projected Hartree--Fock prescription, as already for the total hole--strength
sum rule (\ref{Eq30}), this is only approximately true. However, again the violation
is small.

\begin{figure*}
\begin{center}
\includegraphics[angle=-90,width=12cm]{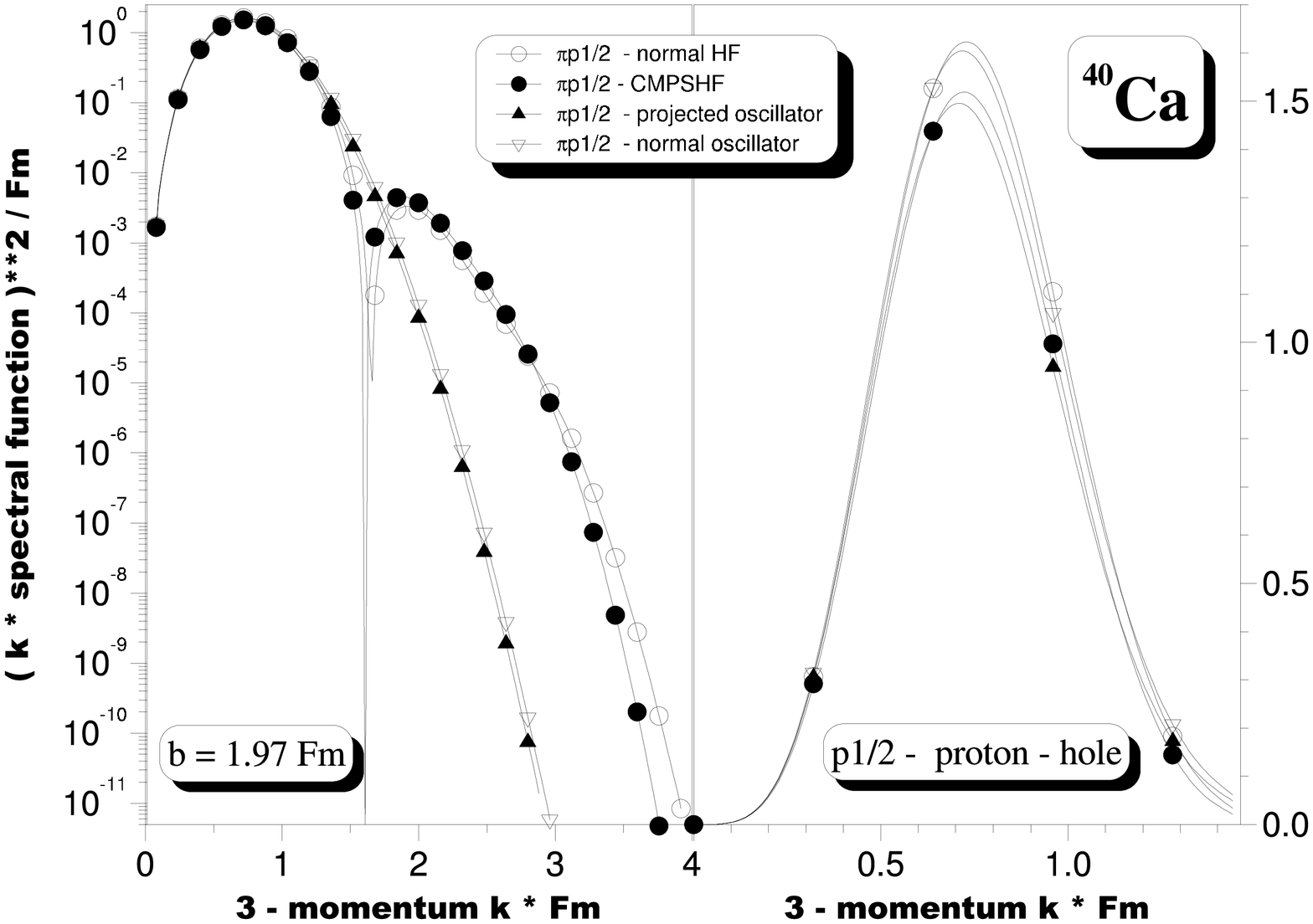}
\caption{
Same as in Fig. 17, but for the p1/2--proton--hole in the nucleus
$^{40}$Ca.
}
\end{center} 
\end{figure*}

\begin{figure*}
\begin{center}
\includegraphics[angle=-90,width=12cm]{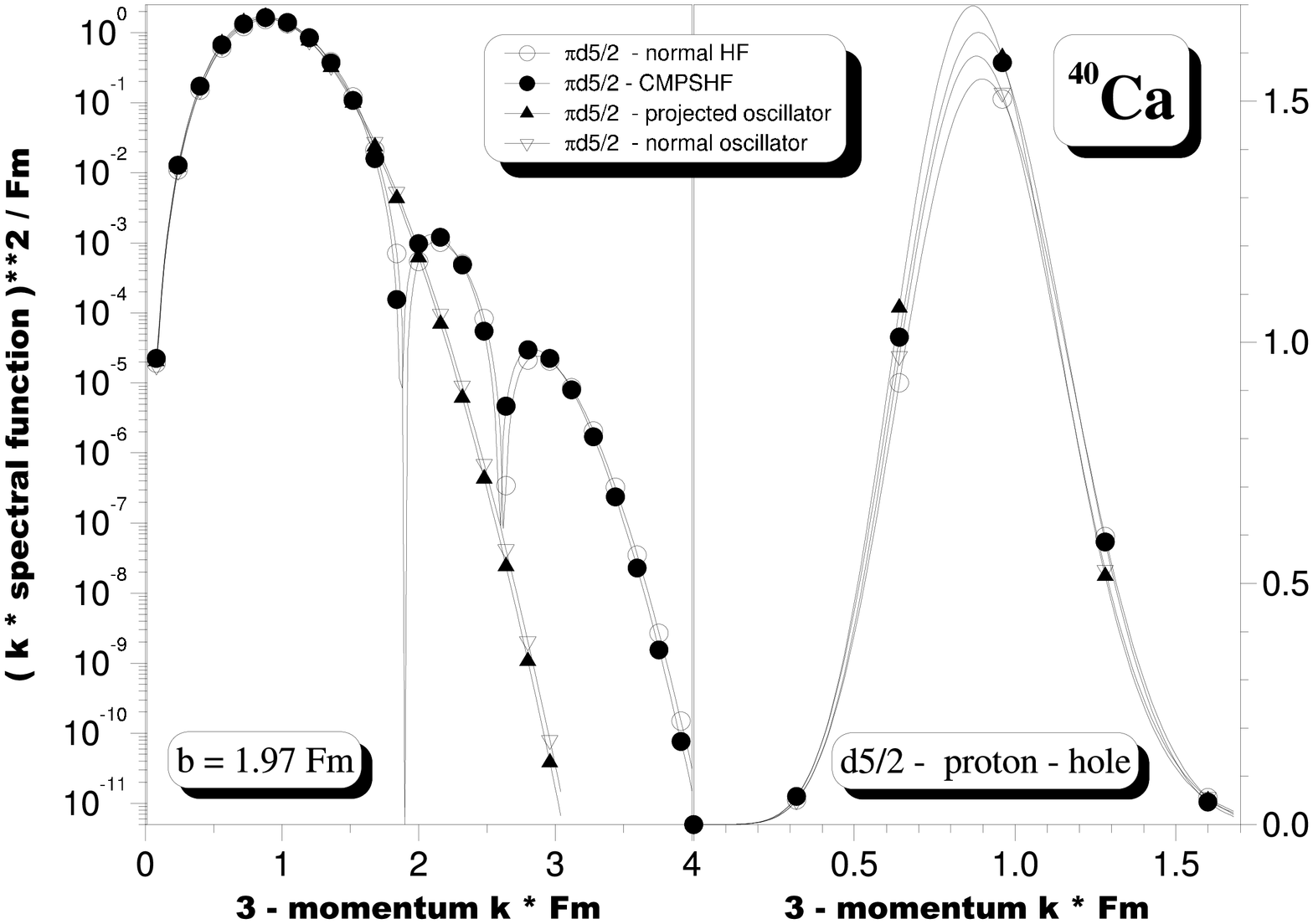}
\caption{
Same as in Fig. 17, but for the d5/2--proton--hole in the nucleus
$^{40}$Ca.
}
\end{center} 
\end{figure*}

\begin{figure*}
\begin{center}
\includegraphics[angle=-90,width=12cm]{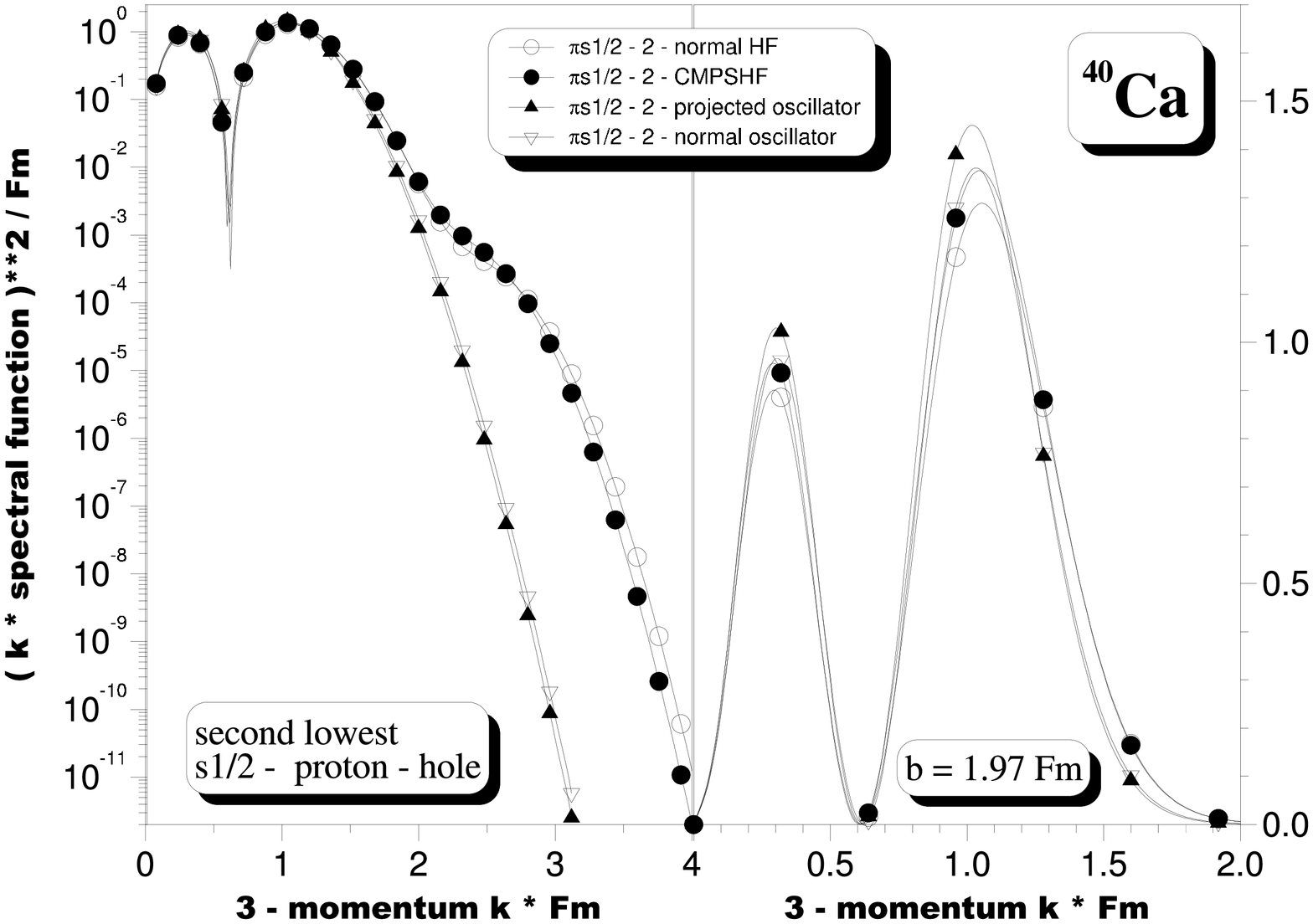}
\caption{
Same as in Fig. 17, but for the second lowest s1/2--proton--hole
in the nucleus $^{40}$Ca.
}
\end{center} 
\end{figure*}

\begin{figure*}
\begin{center}
\includegraphics[angle=-90,width=12cm]{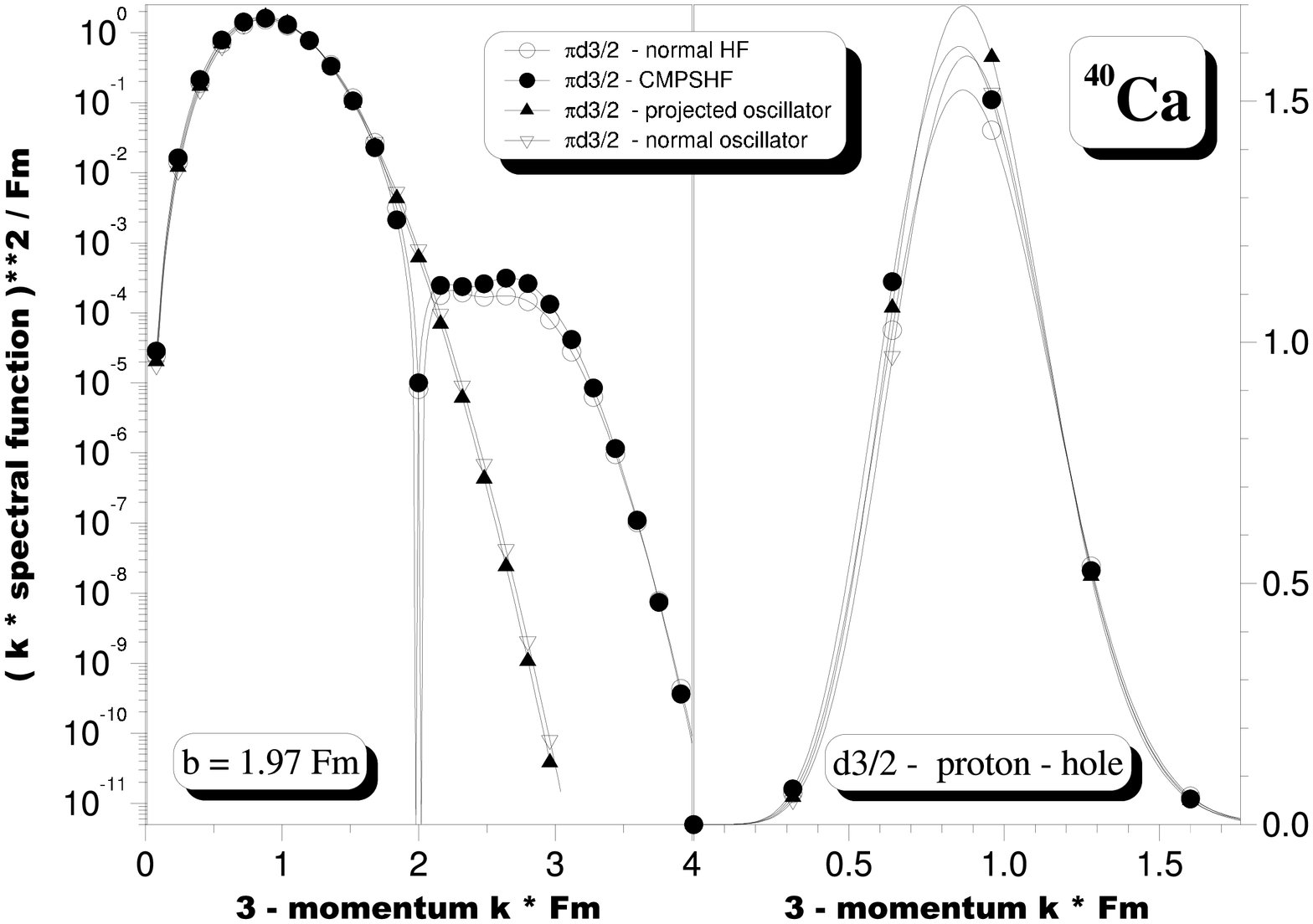}
\caption{
Same as in Fig. 17, but for the d3/2--proton--hole in the nucleus
$^{40}$Ca.
}
\end{center} 
\end{figure*}

The results presented in figure 1 show that our usual picture of an
uncorrelated system has to be changed considerably. This has important
consequences for the analysis of experiments, in which deviations
of the hole spectroscopic factors from one are usually interpreted as
fingerprints of nucleon--nucleon correlations. However, we have shown
that in a Galilei--invariant description the spectroscopic factors
even of an uncorrelated system differ from one considerably. Only
deviations from the projected results out of figure 1 can be related
to non--trivial correlations.

We shall now turn our attention to the reduced spectral functions.
Again only the results for the proton--hole states will be presented.
The results for the different proton--hole states in the various
considered nuclei are presented in figures 2 to 22. The figures
show the square of the reduced spectral functions times the square of the
3--momentum as functions of the 3--momentum. In the left part of each figure
the results are given in a logarithmic, in the right part in a linear
scale. Open circles refer to the normal results (eq. (\ref{Eq7})) for the
ground states obtained with standard spherical Hartree--Fock subtracting
the kinetic energy of the center of mass motion from the Hamiltonian
before the variation in ref. \cite{ref1.}. The integrals over 3--momentum from zero to
infinity yield for these curves always one. Closed circles denote the results
(eq. (\ref{Eq22})) for the ground states obtained by Galilei--invariant projected
Hartree--Fock calculations (CMPSHF) in ref. \cite{ref1.}. Here the integral yields
the projected spectroscopic factors out of eq. (25), which are displayed
by full (or crossed) symbols in figure 1. Open inverted triangles represent
again the results (\ref{Eq7}) of the normal approach, however, now for simple
oscillator ground states. The integrals of these functions are obviously again
all equal to one. Finally, full triangles are used for the (analytically
obtained) projected oscillator results. Here the integrals yield the
harmonic oscillator spectroscopic factors out of eq. (\ref{Eq26}), which are displayed
by open symbols in figure 1.

Figure 2 displays the reduced proton spectral functions for
s1/2--proton--holes in $^4$He. Here oscillator and Hartree--Fock results are
rather similar at low momenta (below about 2 inverse Fm), while at higher
momenta they differ considerably due to the major shell mixing in the latter.
On the other hand the projected results for both approaches differ
considerably from the unprojected ones already at low momenta. Since
(see figure 1) the integrals of all curves give almost the same spectroscopic
factor one, this large difference is entirely due to the fact that the
projected approaches yield relative wave functions instead of the usual ones.
The difference between relative and usual wave functions is obviously largest
in $^4$He and decreases with increasing mass number.

Figure 3 shows the same plots for the s1/2--proton--holes in $^{12}$C.
Again a rather large similarity between oscillator and Hartree--Fock
results is obtained at low momenta, while large differences are seen
above about 2 inverse Fm. Because of the larger mass, the difference of
relative and usual wave functions is here less pronounced, however,
now the projected results are quenched by about 18 percent with respect
to the normal ones due to the considerably smaller spectroscopic factor.

For the p3/2--proton--holes in $^{12}$C in figure 4, the difference
of oscillator and Hartree--Fock results due to the larger major shell mixing
increases already at low momenta. Instead of a quenching, because of the
larger spectroscopic factor here an enhancement of the projected results
with respect to the unprojected ones is seen.

The results for the various proton--hole states in $^{16}$O displayed
in figures 5 to 7 show, as expected, rather similar features as those for
$^{12}$C, except that here the major shell mixing becomes even more
important, so that the deviations of the Hartree--Fock results from
the oscillator ones are larger and start already at lower momenta.
Again, according to the corresponding spectroscopic factors, the
projected s1/2--spectral functions (figure 5) are quenched with respect
to the usual ones, while for the projected p3/2-- (figure 6) and 
p1/2--spectral functions (figure 7) an enhancement is obtained.

Similar arguments hold for the various hole states in $^{28}$Si (figures
8 to 11). Here the projected spectral functions for the s1/2-- (figure 8)
and p3/2--holes (figure 9) are quenched, while those for the d5/2--holes
(figure 11) show an enhancement due to the corresponding spectroscopic
factors. That such an (though small) enhancement is also seen for the
projected p1/2--hole--spectral functions (figure 10) is due to the absence
of the d3/2--state in the ground state and has been discussed already
above.

This pattern is essentially repeated for $^{32}$S (figures 12 to 16). The
projected spectral functions are quenched for the lowest s1/2-- (figure 12)
and the p3/2--states (figure 13), enhanced for the d5/2-- (figure 15) and
second lowest s1/2--states (figure 16), while for the p1/2--state 
(figure 14) though belonging to the second but last occupied shell because
of the occupied d5/2-- and unoccupied d3/2--orbits again a slight enhancement
is obtained.

Finally, for the doubly closed  shell nucleus $^{40}$Ca (figures 17 to 22)
the spectral functions out of the last occupied shell (d5/2, second lowest
s1/2 and d3/2 in figures 20, 21 and 22, respectively)
are enhanced by almost the same factor, while for the holes with excitation
energy $1\hbar\omega$ (p3/2 and p1/2 in figures 18 and 19) or $2\hbar\omega$
(the lowest s1/2 in figure 17) the projected results are quenched with
respect to the normal ones.

Note, that in all cases, though sometimes a little obscured by the logarithmic
plotting, considerable differences are seen between the projected and the
normal Hartree--Fock hole--spectroscopic functions, which can not be explained
by ``quenching'' or ``enhancement'' due to the corresponding spectroscopic
factors alone. This demonstrates that the single particle wave
functions obtained by the Galilei--invariant  Hartree--Fock prescription
are rather different from those obtained via the usual Hartree--Fock
approach as it has been demonstrated already by other observables in ref.
\cite{ref1.}.

\section{Conclusions.}
Normally, we describe the ground state of an uncorrelated A--nucleon
system by a single Slater--determinant, in which the energetically lowest
A single particle states are fully occupied while the higher orbits are
empty. The hole--spectral functions of such a system are then the
Fourier--transforms of the single particle states it is composed of,
and the hole--spectroscopic factors are all equal to one.

This simple picture, however, is not true any more, if Galilei--invariance
is respected. As already demonstrated in ref. [10,2] using simple oscillator
configurations for the ground state of some doubly--closed major shell
nuclei, Galilei--invariance requires a considerable depletion of the
spectroscopic factors for hole--states out of the second and third but last
shell below the Fermi--energy, while those for the hole--states out of the
last shell are enhanced, so that the sum rule for the total hole--strength
remains conserved.

These results are nicely confirmed even for the more realistic Hartree--Fock
wave functions analyzed in the present paper. The Galilei--invariance
respecting hole--spectroscopic factors for the Hartree--Fock ground
states resulting from calculations with projection into the center of mass
rest frame before the variation are almost identical to the projected
oscillator results from ref. [2] and thus fulfill the sum rule for the
total hole--strength in a very good approximation, too. Furthermore,
in both the oscillator as well as the Hartree--Fock description,
Galilei--invariance induces an interesting spin--orbit effect into
the 0p--shell of the closed subshell nuclei $^{28}$Si and $^{32}$S.
On the other hand, as expected because of the major shell mixing, the 
hole spectral functions obtained in the projected Hartree-Fock prescription,
are quite different from the simple projected oscillator ones.

The results clearly show, that not only in the simple oscillator approximation
but also for more realistic approaches the simple picture of an uncorrelated
system has to be changed considerably if Galilei-invariance is respected.
This may have serious consequences for the analysis of correlations
in the nuclei, since the correct uncorrelated reference is considerably
different from that which is usually assumed.

\begin{acknowledgement}
We are grateful that the present study has been supported by the Deutsche 
Forschungsgemeinschaft via the contracts FA26/1 and FA26/2.
\end{acknowledgement}

\end{document}